\documentclass[10pt,preprint]{aastex}
\usepackage[usenames]{color}

\providecommand{\x}{{$\times$}}
\providecommand{\e}{{\tiny E}}

\newcommand{\kms}{ km s$^{-1}$}
\newcommand{\cc}{ cm$^{-3}$}
\newcommand{\lsol}{\hbox{$L_\odot$}}                   % Solar luminosity
\newcommand{\nh}{$N_{\rm{H}}$ }                         % N_H
\shorttitle{}
\shortauthors{}

\begin{document}
\title{Spitzer Observations of Molecular Hydrogen in Interacting Supernova Remnants}

\author{John W. Hewitt\altaffilmark{1},
Jeonghee Rho\altaffilmark{2}, 
Morten Andersen\altaffilmark{2}, and 
William T. Reach\altaffilmark{2}
}

\altaffiltext{1}{Department of Physics and Astronomy, Northwestern University, Evanston, IL 60208}
\altaffiltext{2}{Spitzer Science Center, California Institute of Technology, Pasadena, CA 91125}

\begin{abstract}
With $Spitzer$ IRS we have obtained sensitive low-resolution spectroscopy from 5 to 35\micron\ for
six supernova remnants (SNRs) that show evidence of shocked molecular gas: Kes 69, 3C 396, Kes 17,
G346.6-0.2, G348.5-0.0 and G349.7+0.2. Bright, pure-rotational lines of molecular hydrogen are
detected at the shock front in all remnants, indicative of radiative cooling from shocks interacting
with dense clouds. We find the excitation of H$_2$ S(0)-S(7) lines in these SNRs requires two
non-dissociative shock components: a slow, 10\kms\ C- shock through clumps of density 10$^6$
cm$^{-3}$, and a faster, 40--70\kms\ C- shock through a medium of density 10$^4$ cm$^{-3}$ The
ortho-to-para ratio for H $_2$ in the warm shocked gas is typically found to be much less than the
LTE value, suggesting that these SNRs are propagating into cold quiescent clouds. Additionally a
total of 13 atomic fine-structure transitions of Ar$^+$, Ar$^{++}$, Fe$^+$, Ne$^+ $, Ne$^{++}$,
S$^{++}$, and Si$^{+}$ are detected. The ionic emitting regions are spatially
segregated from the molecular emitting regions  within the IRS slits. The presence of ionic lines
with high appearance potential requires the presence of much  faster, dissociative shocks through a
lower density medium. The IRS slits are sufficiently wide to include regions outside the SNR  which
permits emission from diffuse gas around the remnants to be separated from the shocked emission. We
find the diffuse H$_2$ gas projected outside the SNR is excited to a  temperature of 100--300 K with
a warm gas fraction of 0.5 to 15 percent along the line of sight.

\end{abstract}

\section{Introduction}

As a supernova remnant evolves into its surrounding environment, it
becomes the dominant reservoir of energy that drives dynamics and chemistry in
the interstellar medium \citep{mckee77}. 
As massive stars in general do not drift
far from their parent cloud in their short lifetimes, supernovae near
giant molecular clouds (MCs) should be common. Yet there are
only a handful of SNRs which have been clearly identified as
interacting with MCs. 
Pure-rotational transitions of molecular hydrogen(H$_2$) are
excellent indicators of such interactions as high columns of H$_2$ can
survive the passage of the shock.

Molecular hydrogen is the most abundant molecule in the
Universe, playing a crucial role in the interstellar medium(ISM) as a
catalyst of chemistry and a major coolant. 
H$_2$ is
excited by energetic processes such as shocks and photo-dissociation regions, with numerous
bright emission lines in the near- and mid-infrared(IR) available as
detailed diagnostics of the excited gas.  The low-lying pure rotational
transitions of H$_2$ probe the bulk of the gas involved in the shock
interaction and are excellent diagnostics of the shock conditions. Due
to their relatively  low critical densities the S(0) to S(7)
transitions are not good diagnostics of the excitation mechanism;
however, supernova remnant shocks have the advantage of being
comparatively simple systems in which to understand detailed shock
processes as there is no complicating stellar radiation source.  The
study of Galactic supernova remnants serves as an important template
for other shock interactions, 
for example the spectacular galaxy-scale shocks seen in Stephan's Quintet \citep{appleton06}.

In this paper we report the results of mid--IR long-slit spectroscopy of six
SNRs which are identified as bright in IR molecular lines. These SNRs
have not been previously studied in the mid-IR and represent a
diversification of the common sample of nearby ineracting SNRs such as IC
443, W28, W44 (see Rho et al. 2001 and Reach \& Rho 2005 and references therein). 
The observations are briefly described in $\S$2, followed
by a summary of the properties and observed line emission for each
source in $\S$3. Section \ref{sec:h2ex} pertains to the
physical conditions of the detected pure rotational H$_2$ lines.
Sections \ref{sec:h2diag} and \ref{sec:iondiag} reconciles the observed
molecular and atomic lines, 
respectively. Section \ref{sec:diffuse} presents a
discussion of diffuse H$_2$ observed outside the shock front.

\section{Observations}\label{sec:observations}

Motivated by the detection of eighteen IR-bright SNRs using the GLIMPSE
survey \citep{reach06}, we have selected six SNRs which display {\it Spitzer} Infrared Array Camera (IRAC) colors indicative of molecular shocks. Of the six interacting remnants we
consider in this work, four SNRs -- Kes 69, G346.6-0.2, G348.5-0.0,
and G349.7+0.2 -- have previously identified OH(1720 MHz) masers which
have been established as excellent signposts of interaction with an
adjacent MC \citep{fyz03mmsnrs}. For Kes 17 and 3C 396 the
evidence for interaction was less clear, but both have bright
filamentary radio shells and IRAC band ratios indicative of molecular
shocks \citep{reach06}. 

The {\it Spitzer Space Telescope} Infrared Spectrograph (IRS) was used to obtain
spectra of these SNRs to characterize the nature of emission using the
short-low(SL, 5.2-15\micron) and long-low(LL, 14-42.4\micron) modules and with a
resolving power of R $\approx$ 60-125. In each remnant a single position
coincident with a peak in the IRAC channel 2 emission was chosen (see Table
\ref{tbl:properties}). Exposure times of 150 and 180 seconds were used for the SL
and LL modes. As the exposures were nodded, the effective integration time for the
overlapping region between both orders is doubled typically producing a line
signal-to-noise ratio of 20. For G349.7+0.2 the remnant is bright enough that only
half the integration time was needed. A cluster of reference sky positions devoid
of IR sources were observed within a few degrees of the targets. Subtracting the
median averaged reference position from the source positions successfully
mitigated most rogue pixels. The remaining rogue pixels were removed manually
utilizing the IRS CLEAN program provided by the Spitzer Science Center.
Additionally the slits were nodded with respect to the target position so that
different sections of the slit observed the same overlapping target position.

Spectra were extracted from the region of overlap between the IRS SL and LL slits
and aperture correction was applied. We did not find it not necessary to defringe
the observed spectra. For each target we subtracted a local background using a
nearby position within the same IRS nodded pair. In all cases a position which
lies outside the remnant, and has the lowest background region from one of four
co-added observations is used. As the short-low and long-low slits are aligned
roughly perpendicular to each other, we must use different locations as
representative local backgrounds for the short- and long-orders. In practice the
backgrounds did not differ substantially and produced only minor discontinuities
between the two IRS orders. The H$_2$ S(2) through S(7) lines are only detected
from the SNR interaction regions and are not affected by a local background subtraction.
However, the H$_2$ S(0) and S(1) lines show diffuse emission across the slit, even outside the SNR, but with a clear enhancement of emission at the SNR shell.  The
detailed properties of the remnant and diffuse background are discussed separately
in $\S$\ref{sec:h2ex} and $\S$\ref{sec:diffuse}, respectively.

\section{Results}\label{sec:results}

\subsection{IRS Spectra and Line Brightnesses}

\begin{figure} \centering
\includegraphics[width=6in]{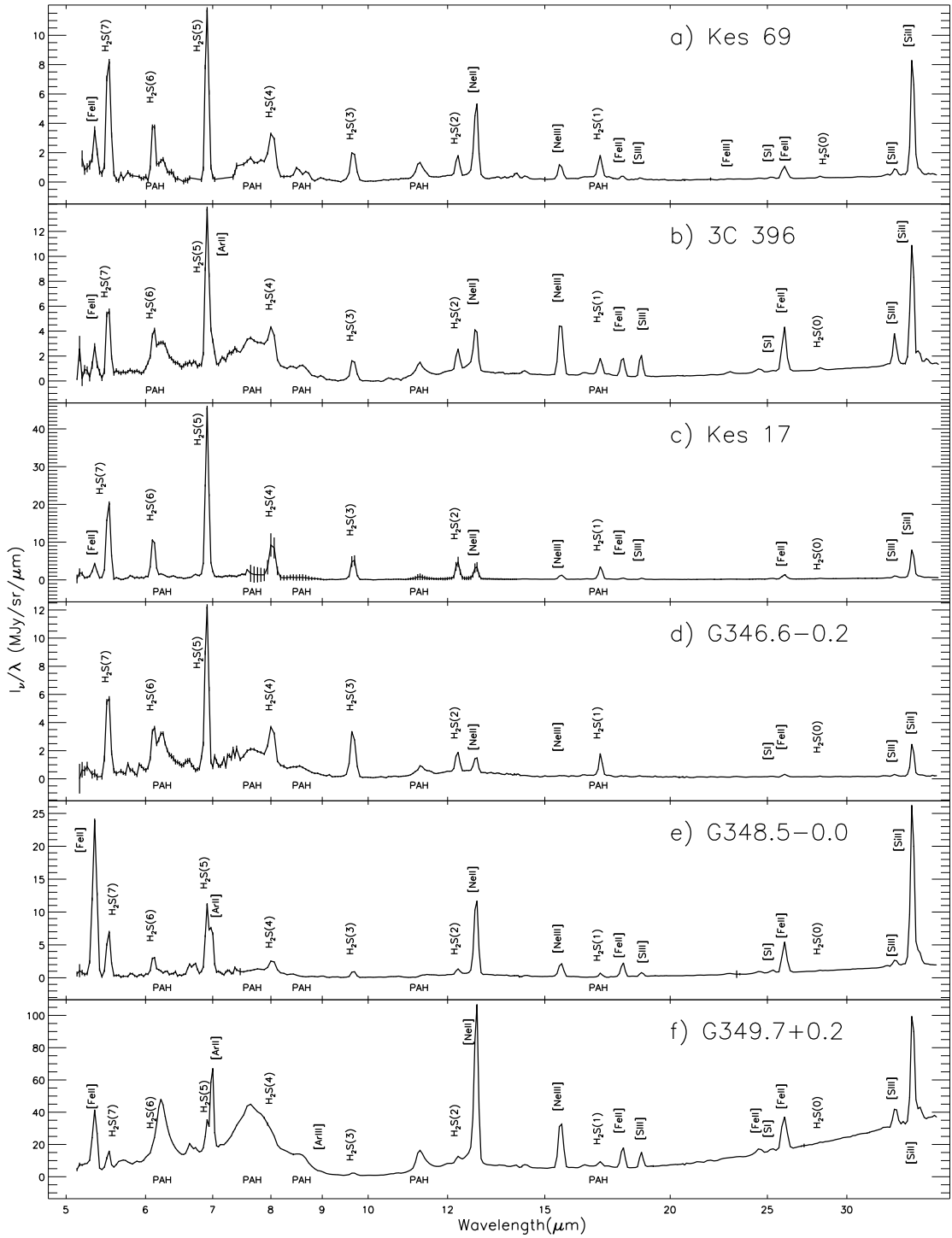}
\caption{Observed IRS spectra after local background subtraction. Detected lines and PAH features
are labeled. From top to bottom: (a) Kes 69, (b) 3C 396, (c) Kes 17,  (d) G346.6-0.2, (e) G348.5-0.0
and (f) G349.7+0.2.
%MA TODO labels are too small, cannot be read. Maybe a less busy y axis? 
\label{fig:spectra}}
\end{figure}

\begin{figure} \centering
\includegraphics[width=3in]{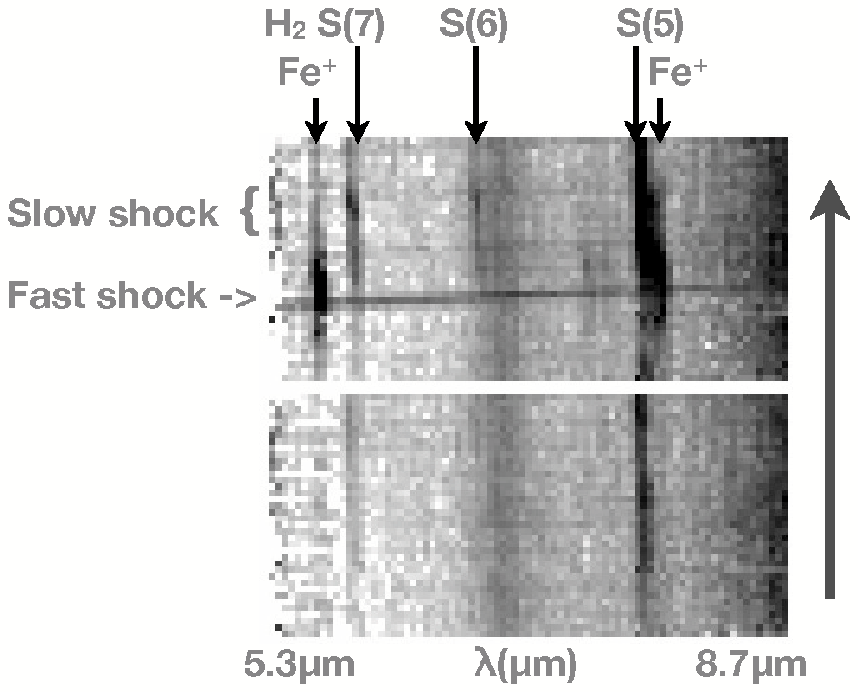}
\includegraphics[width=2.2in]{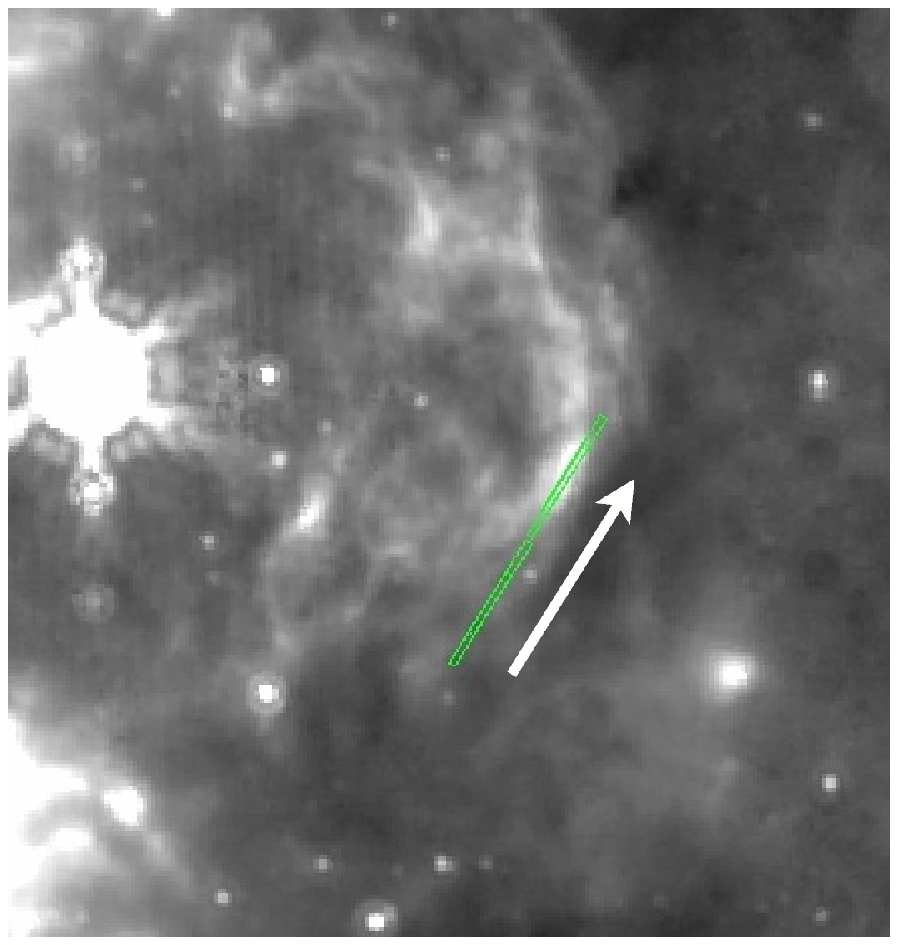}
\caption{Left: Spectral images for the short-low second order of IRS for 3C 396 between 5.1 and
8.7\micron . Right: MIPS 24\micron\ image of SNR 3C 396 with the IRS SL2 slit overlaid. Arrow is
drawn to indicate the direction along the slit. \label{fig:SL2}}
\end{figure}

\begin{figure} \centering
\includegraphics[width=5in]{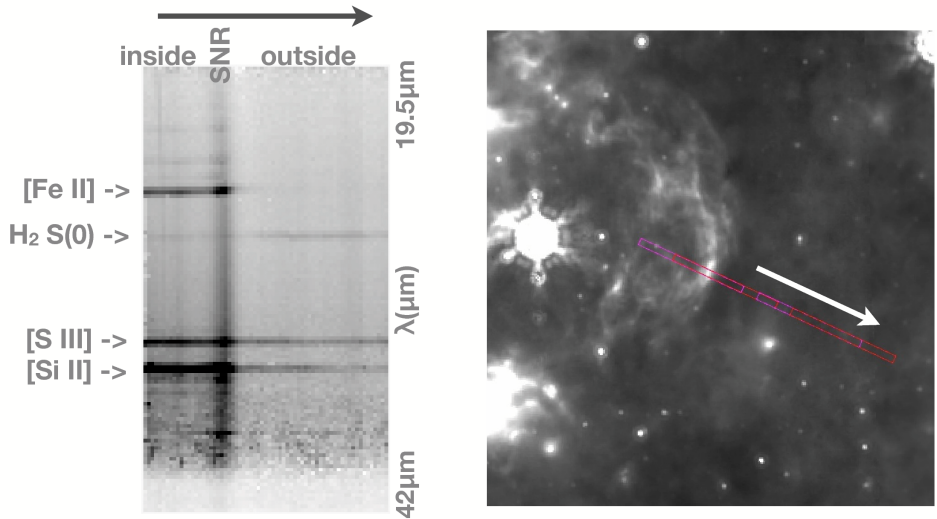}
\caption{Left: Spectral image for the long-low first order of IRS for 3C 396 between 19.5 and
42.4\micron . Note that the spectrum is severely degraded at
wavelengths greater than 35\micron\ toward the bottom of the spectral image.
Right: MIPS 24\micron\ image of SNR 3C 396 with the IRS LL1 slit overlaid. Arrow is
drawn to indicate the direction along the slit. 
%MA TODO 42.4 microns? why not only show the spectrum until 35 microns or at least only 38? 
\label{fig:LL1}}
\end{figure}

IRS spectra of SNRs Kes 69, 3C 396, Kes 17, G346.6-0.2, G348.5-0.0 and
G349.7+0.2 after local background
subtraction are presented in Figure \ref{fig:spectra}
with prominent emission lines and PAH features labeled. 
For all remnants bright pure
rotational transitions of H$_2$ (0,0) S(0) through S(7) are detected. Atomic
sulfur and several ionized fine-structure transitions -- Ar$^+$, Ar$^{++}$,
Fe$^+$, Ne$^+$, Ne$^{++}$, S$^{++}$, and Si$^{+}$ -- are also observed. 
The multiple
molecular and ionic transitions detected with $Spitzer$ are the first clear evidence for
interaction in most of these remnants. For 3C 396 there is evidence that molecular
and ionic lines are spatially segregated within the IRS slits shown in Figures
\ref{fig:SL2} and \ref{fig:LL1}. A steadily rising background continuum level and
in some cases PAH emission are also seen after local background subtraction, but
detailed analysis and discussion of this emission is deferred to a future paper.
In general there is good agreement between the IRS spectra and the nature of IR
emission predicted from the simple color template analysis of IRAC band emission
by \citet{reach06}.

Properties of the SNRs considered in this work are given in Table
\ref{tbl:properties}. The best distance determination is used to obtain a
corresponding physical size of the remnants radio shell. For Kes 69, 3C396 and
G349.2+0.2, we used extinction values from the literature. For Kes 17, G348.5-0.0
and G346.6-0.2, we computed the visual extinction in two or three methods listed
below and  found the most consistent values. First, the IRS spectra are fit by the
lines and continuum (using PAHFIT; Smith et al. 2007)  where the silicate
absorption dips are important  to determine the extinction. We have verified the
extinction solutions of PAHFIT by estimating the depth of the 9.7\micron\ silicate
feature  against the continuum of the remnant. We used the extinction curve of
\citet{draine03} with a visible extinction parameter, R$_V$ = A$_V$/E(B-V), of 3.1 appropriate for the diffuse ISM in the Milky
Way to derive the extinction from the  9.7\micron\ optical depth. Second, the
total hydrogen column density along the line of sight is derived from X-ray
observations for Kes 17 and G348.5-0.0 using archival ASCA data. The detailed
X-ray analysis will be presented in subsequent paper (Rho et al. in preparation).
Third, we have used CO and HI data \citep{dickey90,dame01} to estimate the total
hydrogen column to the SNR. When the kinematics of the SNR are known from masers,
we can estimate the column of atomic and molecular hydrogen by integrating up to
the systemic velocity of the SNR. This gives an upper limit to the hydrogen column
column which is appropriate if the SNR is located at the far distance solution of
the Galactic rotation curve. This was an important constraint for G348.5-0.0 as it
is the only SNR without detected central X-ray emission or an associated pulsar,
and it is believed to lie at the far distance near SNR CTB 37A. Given there is
still some uncertainty in the appropriate de-reddening values for these SNRS we
list the observed line intensities after local background subtraction in Table 2;
the values after extinction correction are given in Table 3.

Using the distance and extinction given in Table \ref{tbl:properties}, the total
luminosity of IR lines observed for Kes 69, 3C 396, Kes 17, G346.6-0.2, G348.5-0.0
and G349.7+0.2 are derived. In all cases the lowest luminosity source was Kes 69
and the brightest was G349.7+0.2. For H$_2$ S(0) to S(7) we find a total
luminosity of 1.4 to 68 \lsol . For ionic lines with ionization potentials less
than 13.6 eV (Fe$^+$, Si$^+$) and greater than 13.6 eV (Ar$^+$, Ar$^{++}$, Ne$^+$,
Ne$^{++}$ and S$^{++}$) we find a total luminosity between 5 to 1212 and 0.8 to
595 \lsol\ respectively. These two groups of ions are considered separately
because they require different physical conditions and previous spectral line
mapping of interacting SNRs has shown the two are not spatially correlated
\citep{neufeld07}. Line luminosities for each species are given in Table
\ref{tbl:lum} for comparison. The [Si II] 34.8\micron\ line is always observed as
a significant cooling line. For Kes 69, Kes 17 and G346.6-0.2 molecular hydrogen
lines are also significant coolants in the observed IRS wavelengths. For 3C396 and
G348.5-0.0 the cooling from the three main ionic lines of Fe, Ne and Si are
comparable to the cooling from H$_2$ lines. The youngest remnant, G349.7+0.2, has
bright ionic lines that are far more luminous than H$_2$. We note that the
luminosities given are for the observed apertures which varies with wavelength.
For the IRS SL and LL modules the apertures are 3.6$\times$ 8\arcsec and
10.5$\times$20\arcsec\ respectively.

\subsection{Individual Supernova Remnants}
Here a brief description of the properties of each SNR is given, noting
the existing evidence for interaction and the individual results of our
$Spitzer$ IRS observations.

{\noindent \bf G21.8-0.2 (Kes 69)}: 
The SNR Kes 69 has a clear Southern radio shell but little radio emission to the
North. IRAC 4.5\micron\ images, that include emission from H$_2$ show a faint
filament along the prominent Southern radio shell. A single OH(1720 MHz) maser
with a velocity of +69.1\kms\ has been detected in the far Northern extent of the
remnant \citep{frail96}. Recent CO and HI observations show Kes 69 is at a
distance of 5.2 kpc with a systemic velocity of +80\kms\ \citep{tian08}. Near this
velocity OH(1720 MHz) emission along the radio shell indicates a molecular shock
\citep{hewitt08}. ROSAT observations give a column of \nh =2.4$\times$10$^{22}$
cm$^{-2}$ \citep{fyz03mmsnrs}.

{\noindent \bf G39.2-0.3 (3C 396)}: 
3C 396 has a bright radio shell and a central X-ray emission. The western shell
appears brighter in both radio and X-ray emission, suggestive of interaction with
a MC. IRAC images in all bands show a bright IR filament along the western edge
with colors indicating an ionic or molecular shock. There is a compact central
X-ray nebula with  a photon index of 1.5 consistent with a central pulsar wind
nebula \citep{olbert03}. There is also significant thermal X-ray emission from the
interior of the SNR not related to the pulsar wind nebula \citep{harrus99}. The
total hydrogen column for the supernova remnant measured by both Chandra and ASCA
is 4.7$\times$10$^{22}$ cm$^{-2}$ \citep{harrus99,olbert03}. Assuming the SNR and
pulsar wind nebula are related, a distance of 6.5 kpc is obtained from the
hydrogen column.

The obtained spectrum of 3C 396 from the western IR filament, shows a wealth of
both H$_2$ and ionic lines of Ar, Fe, Ne, S, and Si. Figure \ref{fig:SL2} shows a clear spatial separation between the H$_2$ and ionic lines along the western
shell. This is clear evidence for spatial variation of physical conditions along
the shock front. In Figure \ref{fig:LL1} diffuse interstellar emission lines of
H$_2$ S(0), [S III] and [Si II] are seen outside the SNR, whereas [Fe II] is only
prominent inside the remnant. These long slit observations hold an advantage over
spectral mapping of nearby remnants in that the emission associated with the SNR
can be clearly distinguished from diffuse emission using local background
subtraction.

{\noindent \bf G304.6+0.1 (Kes 17)}: 
An incomplete radio shell, Kes 17 has prominent IR filaments with colors
indicating bright emission from shocked  H$_2$ \citep{reach06}. The remnant is not
well studied, but HI absorption suggests a distance of at least 9.7 kpc
\citep{caswell75}. Given the absence of clear kinematic signatures we give lower
limits for the distance and size of Kes 17 in Table \ref{tbl:properties}. Archival
ASCA observations show diffuse interior X-ray emission, allowing us to derive a
total hydrogen column of 3.6$\times$10$^{22}$ cm$^{-2}$ toward Kes 17.

{\noindent \bf G346.6-0.2}: 
This remnant has a faint radio shell filled with diffuse thermal X-ray emission,
making it a mixed-morphology SNR. Five OH(1720 MHz) masers are detected along the
Southern shell of the SNR near a velocity of --76\kms ; Zeeman splitting of the
maser line gives a line-of-sight magnetic field of +1.7$\pm$0.1 mG
\citep{koralesky98}. The kinematics implied by the velocity of the associated
maser gives a near/far distance 5.5 or 11 kpc. Archival ASCA GIS observations show
diffuse X-ray emission from the remnants interior which is best fit by a total
hydrogen column of 2.68$^{+2.32}_{-0.68}\times$10$^{22}$ cm$^{-2}$ and a
temperature of kT = 1.36$^{+0.3}_{-0.36}$ eV. The large hydrogen column and small
size of the remnant favor the further distance of 11 kpc, which we adopt here. We
note that MIPS 24\micron\ observations show diffuse emission covers the remnant,
likely due to a foreground HII region to the Northwest.

{\noindent \bf G348.5-0.0}: 
An overlapping radio shell detected adjacent to CTB 37A, the SNR G348.5-0.0
appears to be a kinematically distinct interacting remnant. Both CO and OH maser
observations indicate interaction with a cloud at --20\kms\ for G348.5-0.0,
whereas CTB37A is found at --65\kms\ \citep{frail96,reynoso00}. A line-of-sight
magnetic field strength of --0.8$\pm$0.1 mG is measured for a maser in G348.5-0.0
\citep{brogan00}. The near/far distance ambiguity for both G348.5-0.0 and CTB 37A
are resolved in favor of the far distance by the HI observations, but not with
sufficient resolution to differentiate between the two bright radio sources
\citep{caswell75}. Archival ASCA observations did not allow a determination of the
hydrogen column density towards G348.5-0.0 so using the established kinematic
velocity of --22.0\kms\ we integrated the total CO and HI column densities to
estimate a column of 2.8$\times$10$^{22}$ cm$^{-2}$ towards the remnant.

\begin{figure} \centering
\includegraphics[width=5in]{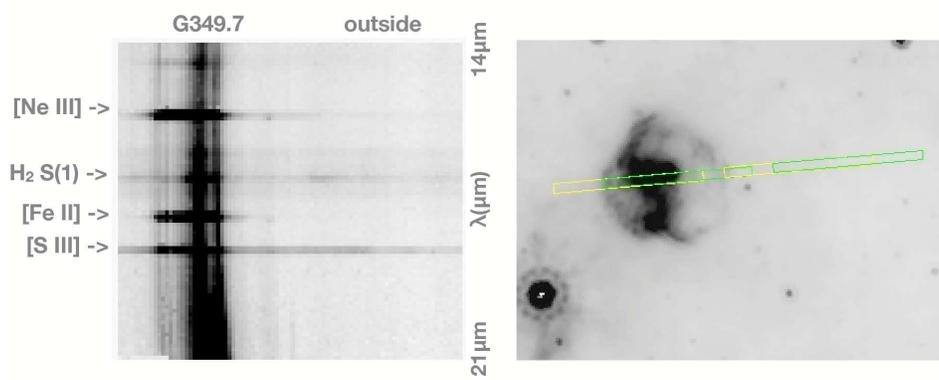}
\caption{Left: Spectral image of the long-low second order of IRS for G349.7+0.2 between 14 and
21\micron . Diffuse PAH emission can be seen around the H$_2$ S(1) line.
Right: MIPS 24\micron\ image of SNR G349.7+0.2 with the IRS LL2 slit overlaid.
\label{fig:g3497_LL2}}
\end{figure}

{\noindent \bf G349.7+0.2}: 
The most luminous and distant SNR within the galaxy with detected OH(1720 MHz)
masers is G349.7+0.2. Numerous molecular transitions indicative of shock
interaction are detected toward including $^{13}$CO, $^{12}$CO, CS, H$_2$,
HCO$^+$, HCN, H$_2$CO and SO\citep{lazendic04oh}.
A kinematic distance to G349.7+0.2 places it at 22 kpc, but rotation curves at
such a large Galacto-centric radius are not well constrained. X-ray observations
yield a neutral hydrogen column of 5.8$\times$10$^{22}$ cm$^{-2}$ and suggest an
age of only 3,500 years \citep{lazendic05}.
\citet{frail96} identified five maser spots at velocities between +14.3 and
+16.9\kms . The impacted cloud appears at $v_{\rm LSR}$=+16.2\kms\ with a linear
size of about 7 pc, a mass of $\sim $104 $M_\odot$ and a volume density of $\sim
$10$^3$ cm$^{-3}$ \citep{dubner04}. The SNR-cloud interaction is thought to be
taking place on the far side of the SNR from a comparison between total absorbing column density and slightly red-shifted wings in the optically thin $^{13}$CO spectra.
Figure \ref{fig:g3497_LL2} shows the IRS LL2 spectra across the remnant. Bright H$_2$ S(1) and [Fe II] lines are seen only associated with the SNR shell. Though some [Ne III] and [S III] is present from diffuse interstellar emission outside the remnant, both lines are clearly enhanced coincident with the bright shell.

\subsection{Excitation of Molecular Hydrogen} \label{sec:h2ex}

\begin{figure} \centering
\includegraphics[width=3in]{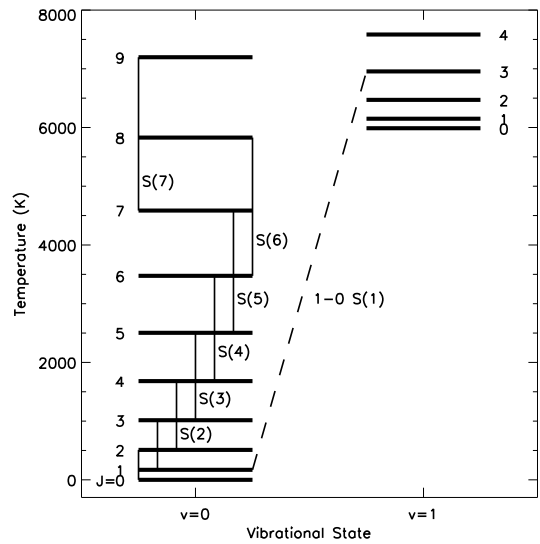}
\caption{Rovibrational energy level diagram for molecular hydrogen with rotational J states labeled.
The S(0) to S(7) transitions available to Spitzer low-resolution IRS observations are indicated with
solid lines. All 'S' transitions are from the upper to lower J state with $\Delta$J=2. The v=1-0
S(1) transition commonly observed at 2.12\micron\ is plotted as a dashed line. Note that the S(7)
transition traces a higher upper excitation state than the 1-0 S(0) and S(1) lines commonly observed
in the NIR.
 \label{fig:h2transitions}} 
\end{figure}

The bright emission from the H$_2$ S(0)-S(7) lines present in all six SNRs is an
excellent diagnostic of the physical conditions of the shocked gas, of which H$_2$
is the dominant constituent. The energy level diagram for H$_2$ is presented in
Figure \ref{fig:h2transitions} with the transitions observed by $Spitzer$ labeled
(by conventional branch notation S indicates $\Delta$J = --2). As these lines are
optically thin they provide a direct measure of the column density for each upper
state J$_u$ = 2 to 9. We note the S(7) transition has a rotational energy higher
than that of the bright H$_2$ 1-0 S(1) line typically observed in the NIR.

\begin{figure} \centering
\includegraphics[width=3.5in]{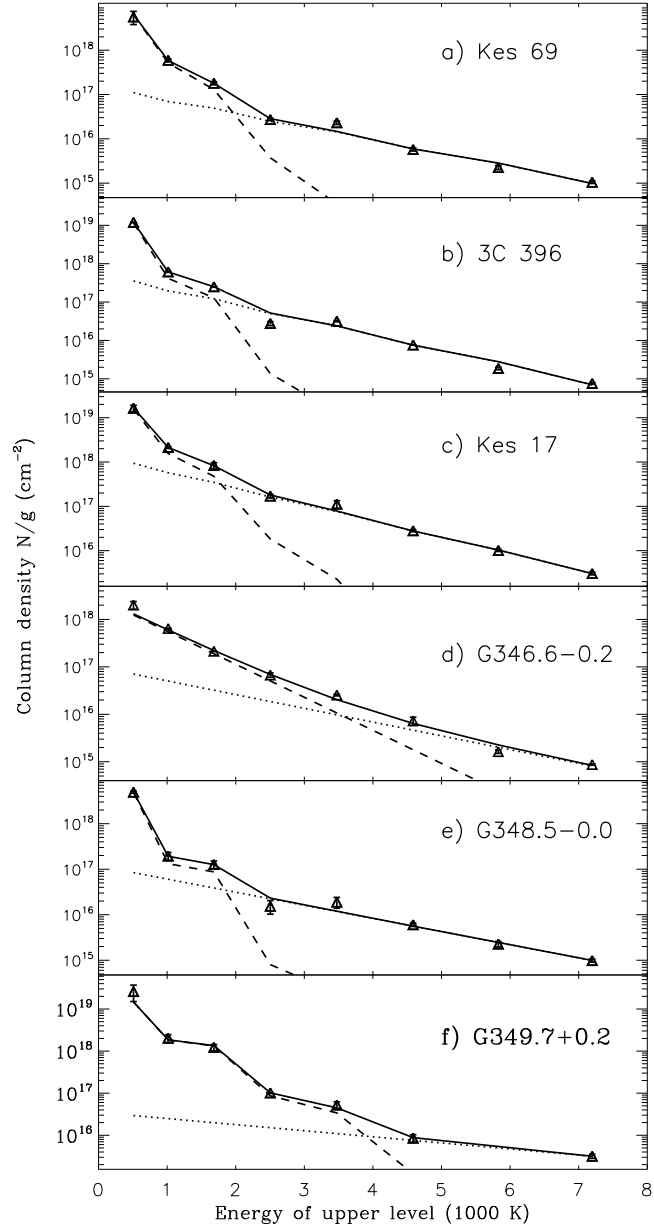}
\caption{Boltzmann diagram for the excitation of H$_2$. From top to bottom: (a) Kes 69, (b) 3C 396,
(c) Kes 17,  (d) G346.6-0.2, (e) G348.5-0.0 and (f) G349.7+0.2. Triangles mark the column density of
the upper J state obtained from the de-reddened brightness with error bars. The best-fit
two-component model is overplotted as a solid dark line. A dashed line shows the contribution of the
warm component; a dotted line shows the contribution of the hotter component. Parameters for the
fits are given in Table \ref{tbl:h2exfits}.
\label{fig:boltz}}
\end{figure}

Rotational diagrams from dereddened H$_2$ line intensities are presented in Figure
\ref{fig:boltz}. Here the column density per magnetic substate is plotted against
the upper energy level, such that a single temperature distribution in LTE follows
a straight line. It is clear that a simple LTE excitation model for the observed
level populations cannot fit both the positive curvature resulting from a range of
gas temperatures, or the zig-zag pattern between subsequent ortho- and para-states
out of equilibrium. However, it is instructive to examine the properties of a two
temperature model which can be well fit to the data. Table \ref{tbl:h2exfits}
gives the best fits from least squares fitting assuming two temperature distributions and simple LTE excitation of H$_2$. Common for the fits are a warm
component with T$_w$ $\sim$ 250--600 K and a column N$_w$ $\sim$ 10$^{20}$
cm$^{-2}$, and a hot component with T$_h$ $\sim$ 1000--2000 K and a column N$_w$
$\sim$ 10$^{19}$ cm$^{-2}$.  In Figure \ref{fig:boltz} the warm component is drawn
as a dashed line, the hotter component is drawn as a dotted line and the total
contribution as a solid line. For the hotter gas component (dotted line) the ortho
to para ratio (OPR) is found to be at or near the statistical equilibrium value of
three. However, the warm component (dashed line) shows a range of OPR between 0.4
and 3, manifested as zig-zags between ortho- and para- levels in the energy
diagrams. This large range in OPR$_w$ was also found for the interacting SNRs 3C391,
IC443, W28 and W44 \citep{neufeld07}. The conversion of ortho to para in shocks is
strongly dependent on temperature, with threshold temperatures at 700 K to start
conversion and 1300 K to rapidly reach the equilibrium value \citep{timmermann98}.
The temperature of warm gas is substantially less than 700 K for all but SNR
G346.6-0.2 which has an OPR$_w$ of 3. For the other remnants an OPR between 0.4
and 1.4 is characteristic of an equilibrium temperature between 50 and 100 K. For
shock velocities $\la$ 20\kms\ the H$_2$ gas will retain the OPR of the pre-shock
gas, with only a small dependence on gas density \citep{wilgenbus00}. Low OPR
values indicate that there has not been substantial pre-heating of the pre-shock
material with which these SNRs are now interacting.

\section{Shock Models}

At the position of the bright IR peak observed in each SNR both strong molecular
and ionized species are present. This discordant mix of spectral lines cannot be
reconciled with a single shock into a uniform medium. We consider two general
classes of shock models. Jump(J) shocks have a discontinuous change of
hydrodyanmic variables, with heating over a negligible thickness sufficient to
dissociate molecules \citep{hm89}. It is possible for H$_2$ and other molecules to
reform in the cooling post-shock region. In contrast continuous(C) shocks have a
transition region in which friction between ions and neutrals occurs over a
sufficiently large thickness that the transition from pre- to post-shock is
continuous and nondissociative for molecular gas of modest ionization
\citep{draine83}. In the following sections we consider the physical parameters of
shocks which can explain the observed IR line emission.

\subsection{H$_2$ Rotational Lines\label{sec:h2diag}}

For each remnant the observed H$_2$ columns are compared to published C- and
J-shock models. The H$_2$ line intensities, particularly at low excitation
temperatures, are typically more than an order of magnitude brighter than
predicted by J-shock models (eg., Hollenbach \& McKee 1989). A comparison with
C-shock models (eg., Draine et al. 1983) shows lines of comparable magnitude, but
a single C-shock could not simultaneously account for the strengths of all observed 
H$_2$ lines.

Using a grid of shock models a detailed fit to the excitation of H$_2$ lines was
performed using least squares fitting. The relevant shock variables are shock
velocity v$_s$ and pre-shock density n$_o$. Two recent C-shock models were used.
The first model of \citet{wilgenbus00} accounts for the non-equilibrium OPR
observed in the warm H$_2$ gas. A grid of computed models spans (log$_{10}$
n$_o$)=3,4,5,6 cm$^{-3}$, v$_s$=10,15,20,25,30,40\kms\ and OPR=0.01,1,2,3. For
shock velocities above a certain critical velocity H$_2$ is dissociated; without
H$_2$ cooling a continuous transition cannot be sustained and the shock becomes
J-type. \citet{lebourlot02} found that C-shocks could be sustained at much higher
velocities than previously thought, ranging up to 23 to 80\kms\ for densities of
10$^7$ to 10$^3$ cm$^{-3}$, with increments matching those of \citet{wilgenbus00}.
The second shock model from \citet{lebourlot02} correctly accounts for shock
speeds up to the cirtical velocity but assumes OPR=3. However, for shock
velocities much less than the critical velocity ($\sim$10-20\kms) there is good
agreement between the work of \citet{lebourlot02} and \citet{wilgenbus00} for
OPR=3. Together these two sets of models form a grid of C-shock models that span
the observed H$_2$ excitation parameters found in Section \ref{sec:h2ex}.

In our model fitting each shock component is given a free linear scaling to
account for uncertainty in the geometry and area of the shock relative to the IRS
aperture. This factor $\Phi_s$ is the ratio of the projected beam size to the
surface area of the shock. If the beam is not completely covered by shocked
material the value of $\Phi_s$ can be less than unity.

\begin{figure} \centering
\includegraphics[width=3.5in]{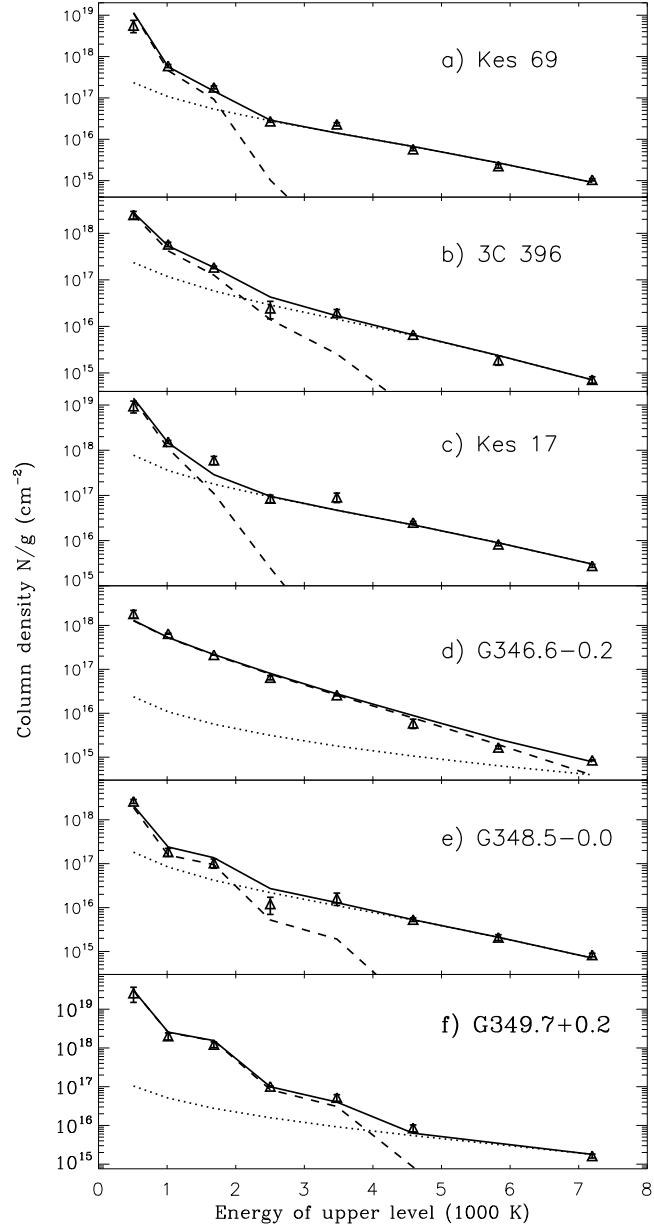}
\caption{Two C-shocks are fitted to the observed excitation of H$_2$. From top to bottom: (a) Kes
69, (b) 3C 396, (c) Kes 17,  (d) G346.6-0.2, (e) G348.5-0.0 and (f) G349.7+0.2. The total of two
fitted shock components is overplotted as a solid line A dashed/dotted line shows the contribution
of the slower/faster shock. Parameters for the fits are given in Table \ref{tbl:shockmodels}.
\label{fig:cshockmodels}}
\end{figure}

No single shock model is found that can reproduce the positive curvature observed
in the Boltzmann diagrams. For a good fit to the data a combination of two shock
models is required. A slow C-shock gives a good fit to the lower excitation lines
(eg., H$_2$ S(1)) but under-predicted the observed brightness of higher excitation
lines (eg., H$_2$ S(7)). In addition to C-shocks we included the J-shock models of
\citet{hm89} as a discontinuous grid of models with densities of 10$^3$-10$^6$
cm$^{-3}$ and shock velocities of 30-150\kms\ in 10\kms\ increments. The results
using a single C-shock with either a second faster C-shock or a fast J-shock are
given in Tables \ref{tbl:shockmodels} and \ref{tbl:cjshockmodels}. Figure \ref{fig:cshockmodels} shows the best
fits to the observed excitation of each SNR which are generally found to be two
C-shocks.

\begin{figure} \centering
\includegraphics[width=3.5in,angle=90]{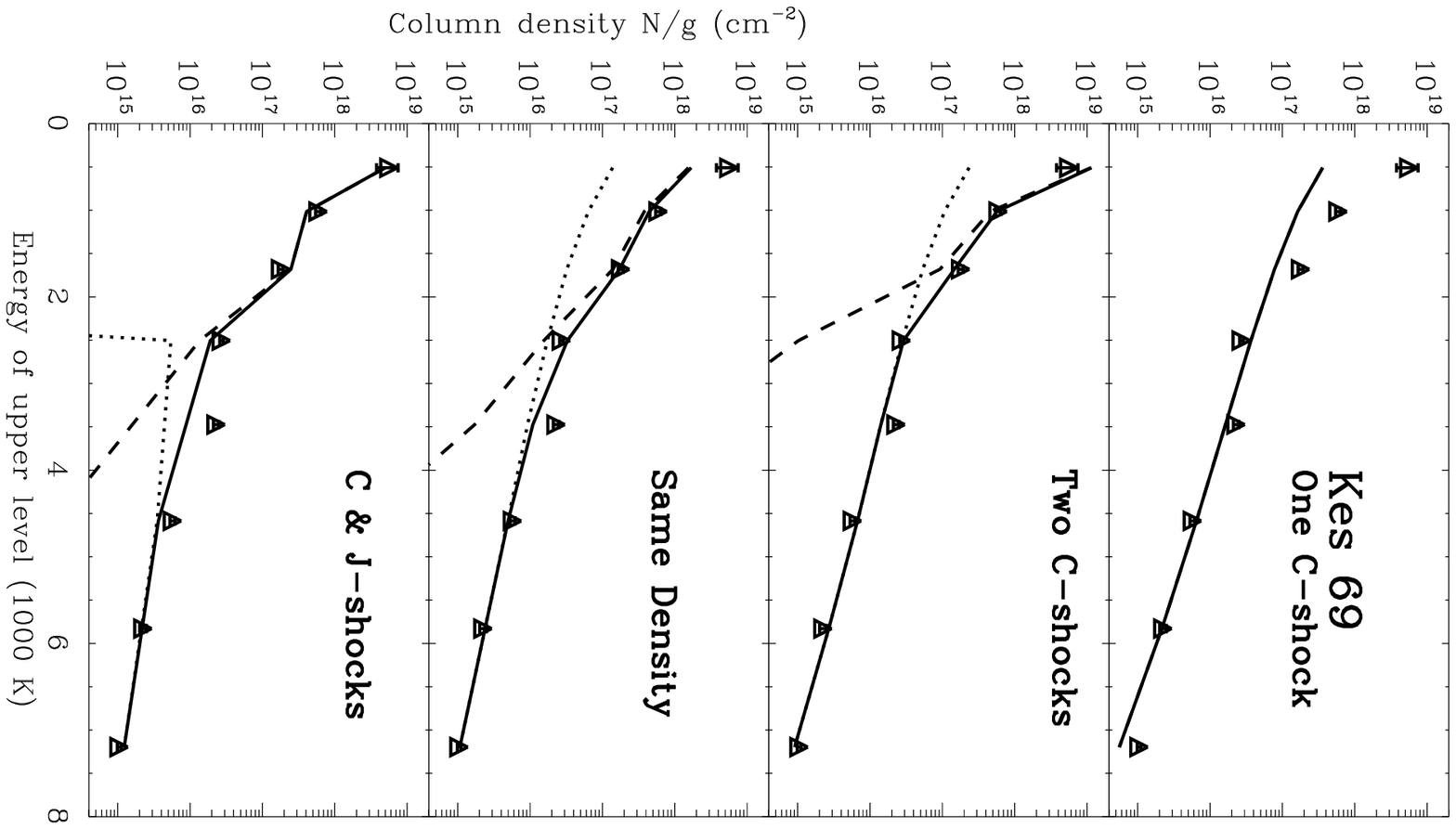}
\includegraphics[width=3.5in,angle=90]{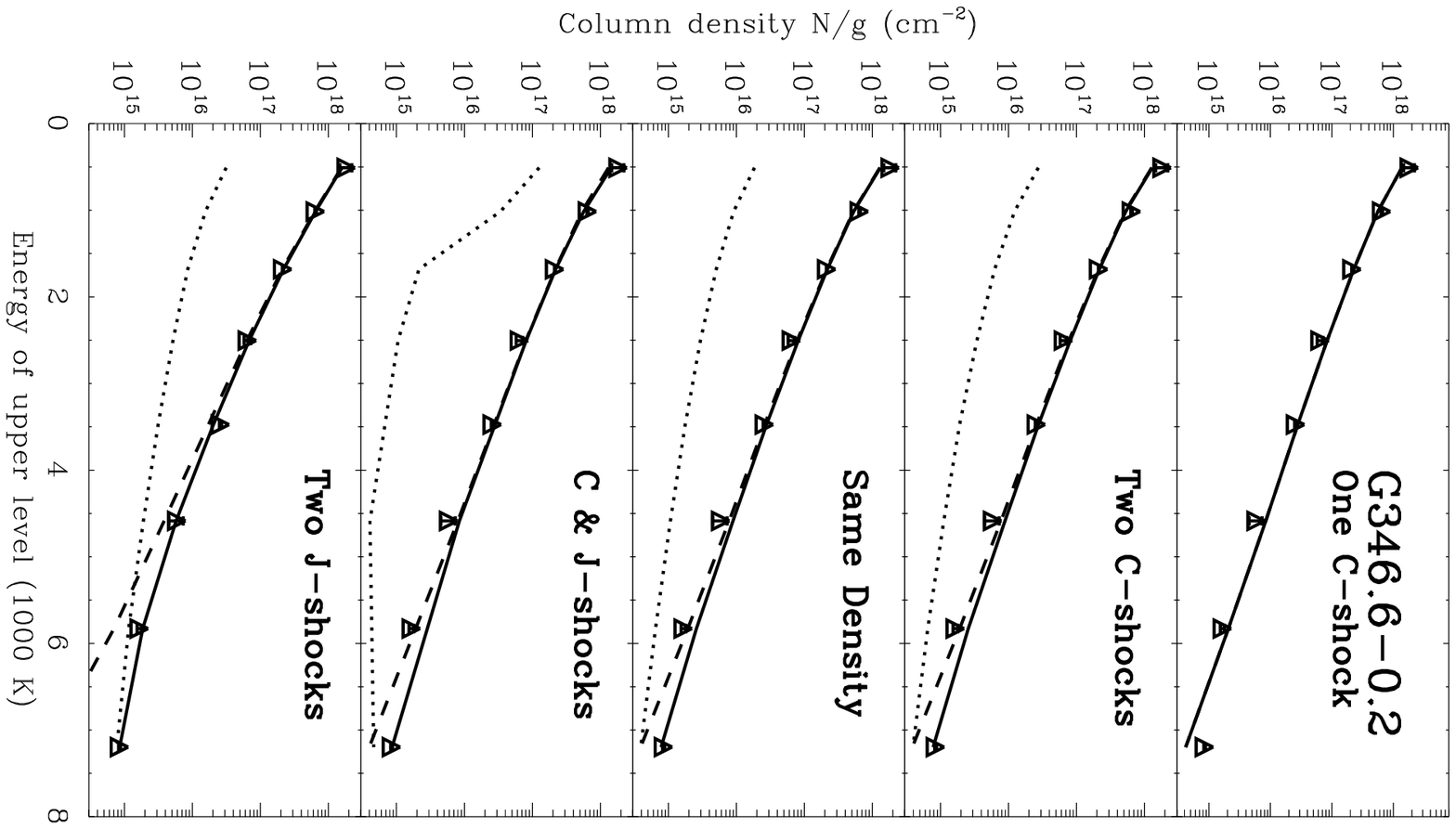}
\includegraphics[width=3.5in,angle=90]{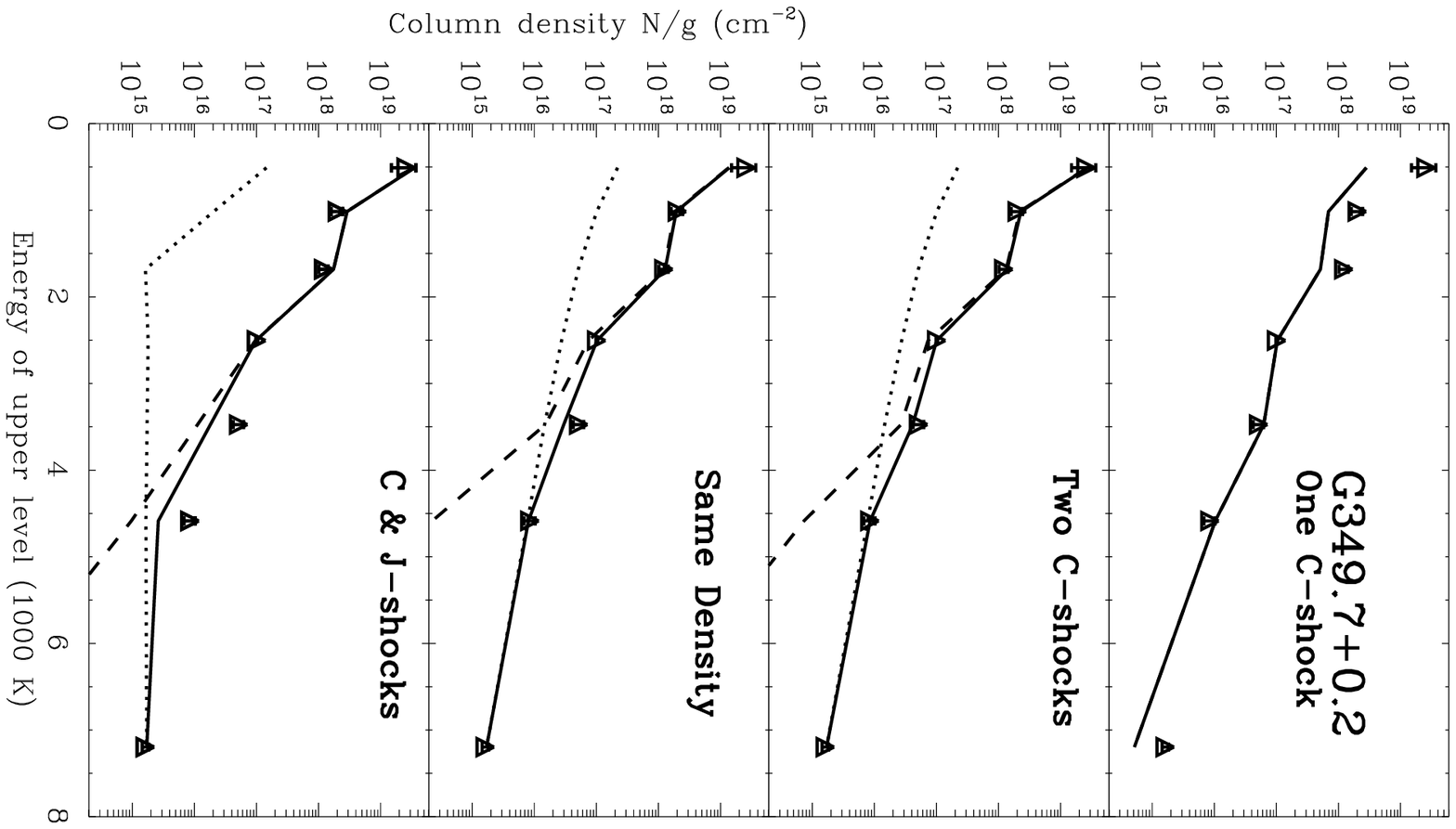}
\caption{A comparison of fitted shock models to excitation of H$_2$ by a combination of shocks for Kes 69 [left], G346.6-0.2 [center] and G349.7+0.2 [right]. From top to bottom: a single C-shock, two C-shocks with different densities and velocities, two C-shocks with the same density of 10$^4$ cm$^{-3}$, a combination of one C-shock and one J-shock. Table \ref{table:extra} has the fitted values and reduced chi-squared values. In cases of multiple shocks the slower shock is plotted with a dashed line and the faster shock with a dotted line. The total contribution of the two shock components is plotted as a solid line.
\label{fig:compare}}
\end{figure}

We find a slow $\sim$10\kms , dense $\sim$10$^6$ cm$^{-3}$ C-shock gives an
excellent fit to the S(0)-S(3) lines, including non-LTE OPR. The higher excitation
lines are well fit by a fast C- or J-shock through a lower density medium
(n$_o\sim$10$^{4}$ cm$^{-3}$, v$_s\sim$30--60\kms ) in which the OPR is quickly
brought to equilibrium. In general, a fast C-shock gives a better fit to the
observed excitation of H$_2$ than a fast J-shock.
The only exception is G346.6-0.2, in which a single C-shock at a velocity of
$\sim$15\kms into a density of 10$^{5-6}$ cm$^{-3}$ and an OPR of 3 reproduces the
observed H$_2$ S(0)-S(6) lines but underestimates the H$_2$ S(7) line. Either a
fast C-shock or a fast J-shock into a lower density medium can produce the S(7)
line but this shock component is not well constrained so either solution is
viable. Figure \ref{fig:compare} shows the alternative J-shock solution given in Table \ref{table:extra}. For
all other remnants a C- and J-shock combination do not do a good job or
reproducing the observed excitation, however we do list the best-fit parameters in Table 7 for comparison with the two C-shock models.
It is possible that our best fit of two C-shock models is a characteristic of 
temporal evolution of an MHD shock \citep{chieze98,cesarsky99}.

We compare the shock fits to a simple analytic treatment of energy conservation in C-shocks \citep[Appendix B]{neufeld06} in order to check the validity of fitting results and to understand associated
physical processes. The characteristic gas temperature of a C-type shock is given by
\begin{equation}
{\rm T_s = 375\ {\rm b}^{-0.36} (v_s/10\ {\rm km\ s}^{-1} )^{1.35}\ K,}
\label{eqn:Ts}
\end{equation}
where b = B/$\mu$G (n$_o$/cm$^{-3}$)$^{-0.5}$. Given the standard assumption that b=1 the warm and hot gas components discussed in Section \ref{sec:h2ex} typically correspond to shock velocities of $\sim$10 and $\sim$30\kms , respectively. This is in general agreement with the parameters fit in the modeling, however the hot gas shock velocity is typically found to be somewhat higher.
The observed column density of shocked H$_2$ is given by 
\begin{equation}
{\rm N_s=4\times 10^{\rm 20}\ b\ n_{04}^{0.5}\ (v_s/10\ {\rm km\ s})^{-0.75}\ \Phi_s,}
\label{eqn:Ns}
\end{equation}
where the preshock density is 10$^4 n_{04}$ cm$^{-3}$ and $\Phi_s$ is the ratio of the projected beam size to the surface area of the shock as before. Even though the preshock density, n$_o$ is not well constrained by the low-J rotational transitions, a comparison between the H$_2$ column for the warm and hot component in Table \ref{tbl:h2exfits} and the expected value of N$_s$/$\Phi_s$ from Equation \ref{eqn:Ns} can establish whether the fitted $\Phi_s$ is reasonable. Using the best-fit model preshock density and shock velocity in Table \ref{tbl:shockmodels} we find the expected values of $\Phi_s$ match the model-fitted $\Phi_s$ within a factor of a few. 
Typically $\Phi_s$ is found to be 0.1--0.5 and 0.03--0.02 for the warm-slow and hot-fast shocks, respectively. 
For Herbig-Haro objects there is significant substructure not resolved by IRS and values of $\Phi_s$ for the warm and hot components are typically 0.2$n_{04}^{0.5}$ and 0.05$n_{04}^{0.5}$ \citep{noriegacrespo02,neufeld06}. Supernova remnants also appear to have significant substructure which IRS observations are not able to resolve.

We now compare the shock models of that fit of our H$_2$  lines with other models and observations.
Le Bourlot et al (2002) applied C-shock models to the H$_2$ lines from Orion and found a best fit with n$_o$=10$^4$ cm$^{-3}$
and two different shock velocities of 40 and 60 km s$^{-2}$. Table \ref{table:extra} shows that two different density models yield a better fits than two different velocities for a single density.      
\citet{burton92} suggested a density of 10$^6$ cm$^{-3}$ with a very low velocity of 10 km s$^{-1}$; 
such a model cannot explain all the higher-J observed H$_2$ emission (see the dashed and dotted lines in Figure 
\ref{fig:cshockmodels}).
\citet{richter95} suggests partially dissociated H$_2$, from a J-shock, but the implied high velocities could not be explained. 
Fits to ISOCAM observations of IC 443 show a density of 10$^4$ cm$^{-3}$ with excitation temperatures of 330-657 K and 1115-1300 K \citep{cesarsky99}.
Our fit results show two main differences from those of Cesarsky et al. (1999); one is a higher density component and the other is
departure from LTE OPRs. The main differences may arise from our inclusion of
S(0) and S(1) lines which could not be observed by ISO. The departure from LTE is independent of shock models and implied by the two temperature fitting of H$_2$ excitation diagram.     
The inferred OPRs are out of equilibrium indicating the time-scale of the shock passage was too short to establish an equilibrium ortho/para ratio \citep{pdf00}.
In the course of shock passage,  temporal non-equilibrium could be due to dynamical processes with colder gas, or fast conversion of ortho-H$_2$ to para-H$_2$ during the formation on the surface of grains. 
Molecular hydrogen formation is mainly governed by the properties of grain surfaces
and shocked medium quickly accreted in the diffuse ISM in even a modest velocity ($\sim$30 km s$^{-1}$) shock \citep{jones94,cazaux02}.
A combination of two C-shock models indicates a shock into very dense (10$^{5-6}$ cm$^{-3}$) gas with 
a low (10-15 km s$^{-1}$) shock velocity 
where sputtering and dust destruction may not be as efficient. 

Given a multi-phase ISM a combination of pre-shock densities is possible. 
A moderate density (10$^3$-10$^4$ cm$^{-3}$), faster shock component has been resolved in W44 and W28 \citep{reach00}.
Widths of H$_2$ in the shock front are measured to be  $\sim$10$^{16}$ cm, consistent with nondissociative shocks into gas with densities of $\sim$10$^4$ cm$^{-3}$. These densities are consistent with our fast C-shock component of 40-70
km s$^{-1}$ (see Table \ref{tbl:shockmodels}).  
Perhaps surprisingly the slower, denser shock has a larger filling factor than this faster, lower density shock in our fitting. At such high densities of 10$^5$-10$^6$ cm$^{-3}$ cooling will be faster with a lower peak temperature and a larger shock thickness if the two shocks are roughly isobaric. Cooling could be significant from molecular lines other than H$_2$ (such as H$_2$O) though the current observations are not sufficient 
to detect such lines, for which observations with a higher resolution spectroscopy are required.

\subsection{Ionic Lines \label{sec:iondiag}}

Strong ionic lines observed toward all SNRs and cannot be reproduced by
steady-state C-shock models (see Draine et al. 1983). We use the J-shock models to
derive the shock velocity and pre-shock density purely from consideration of the
surface brightness of observed ionic lines. Here our purpose is to determine the
gross properties of the fast J-shock responsible for the observed fine-structure
lines\footnote{The [Fe II] 26\micron\ is prominently detected but the spectral
resolution of IRS is not sufficient to distinguish between the [O IV] 25.9 and [Fe
II] 26\micron\ lines. Observations of interacting remnants with ISO and IRS at high
spectral resolution are able to separate the two lines, and find that the [O IV]
25.9\micron\ line does not contribute significantly if at all to [Fe II]
26\micron\ line flux \citep{reach99,neufeld07}. Furthermore the [O IV]
25.89\micron\ line has an ionization potential of 54.9 eV and is only detectable
for very fast shocks in excess of those observed here. We expect that the [Fe II]
25.98\micron\ line is by far dominant for most if not all of the SNRs we consider
here because other transitions of [Fe II] lines are detected.}.

\begin{figure} \centering
\includegraphics[width=4in,angle=90]{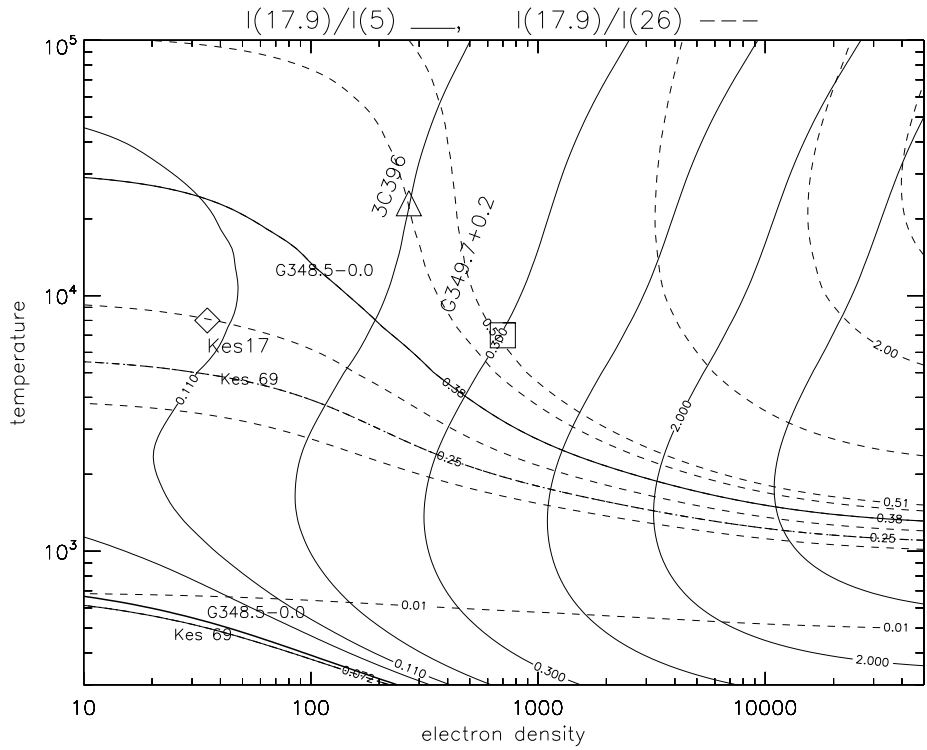}
\caption{Contour plots of of [Fe II] diagnostic line ratios using 5.3, 17.9 and 26$\mu$m lines. The 
observed ratios of G349.7+0.2 (square), 3C 396 (triangle) and Kes 17 are marked. The two ratios for 
Kes 69 (two thick dotted lines) and G348.5-0.0 (two thick solid lines)  do not converge to the same 
temperature and density ranges which may be due to uncertainties of collisional strengths at low 
density and temperature or presence of multi-temperature components. 
\label{fig:feratio}}
\end{figure}

Diagnostics of the [Fe II] lines use excitation rate equations which are presented
in Rho et al. (2001). Figure \ref{fig:feratio} shows contours of line ratios of
17.9/5.35$\mu$m and 17.9/26$\mu$m; the former ratio is mainly sensitive to density
and the latter ratio changes depending on both density and temperature. Updated
atomic data \citep{ramsbottom07} affected (at the $<$10\% level) the contours at low densities
($<$200 cm$^{-3}$). The estimated density and temperatures from [Fe II] ratios are
summarized in Table \ref{table:diagnostics}. A set of electron density and
temperature solutions are possible for each of G349.7+0.2 and 3C 396: 
(n$_e$, T) = (700 cm$^{-3}$, 7000 K) and (270, 2.3$\times 10^4$), respectively. The
ratios of Kes 17 indicates (n$_e$, T) = (100, 7000) showing a low electron density. For Kes
69 and G348.5-0.0, the two ratios did not converge;
the ratio of [Fe II] 17.9/5.35$\mu$m indicates a higher
temperature than that from 
the ratio of [Fe II] 17.9/26$\mu$m. This may be due to
either uncertainties of collisional strengths at low density and temperature or
may indicate presence of multi-temperature gas. 
The ratio of [Fe
II] 17.9/5.35 $\mu$m indicates a low density ($<$ 400 cm$^{-3}$) and low ($<$2000 K) 
temperature. Temperature could not be well-constrained; lines of higher upper
energy levels such as the [Fe II] 1.64$\mu$m line are required to constrain
temperature.

The pre-shock medium density n$_o$ and shock velocity v$_s$ are also constrained
using the strength of the  [Ne II] 12.8, [Si II] 34.8, [Fe II] 26\micron\ lines
(called the three main ionic lines hereafter, see Figure 6 of Rho et al. 2001
which were generated based on Hartigan et al. 1987 and McKee et al. 1984). 
The inferred medium densities and shock velocities are summarized
in Table \ref{table:diagnostics}.
Iron
lines have a high ionization potential and may not arise from the same physical
conditions as these ionic lines with lower ionization potentials. 

\begin{figure} \centering
\includegraphics[width=4in,angle=90]{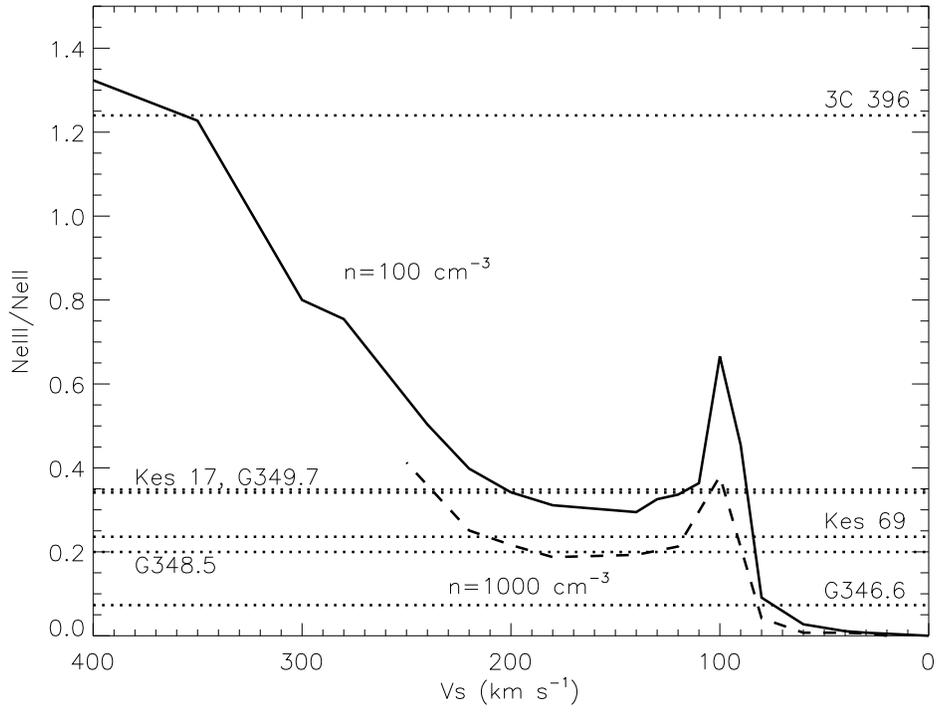}
\caption{Predictions from the models of Hartigan et al. (1987) of the [Ne III] 15.5/[Ne II]
12.8\micron\ ratio as a function of shock velocity for densities of 100 and 1000 \cc . The observed
values for each remnant are marked as dashed lines.
\label{fig:neratio}}
\end{figure}

The ratio of [Ne III] 15.5/[Ne II] 12.8 is also useful to determine shock velocity
and is presented in Figure \ref{fig:neratio} with results summarized in Table
\ref{table:diagnostics}. The inferred shock velocities from Ne ratios are
generally consistent with those from the brightnesses of three  main lines except 3C 396. The only remnant for which the Ne III/Ne II ratio is greater
than unity is 3C 396, suggesting a shock in excess of 300 km/s while  the inferred shock
velocity from the line intensities of [Si II],[Fe II], [Ne II] are 100-150 km
s$^{-1}$. 
The [Ne III] in 3C 396 may arise from fast shocks than the lower-ionization state lines.

\subsection{ Molecular and Ionic Shocks and Ejecta Contribution} 

Estimates of the shock ram pressure (P$_s$) can be derived from the shock modeling in the previous two sections. The best-fit parameters from both C-shocks derived from H$_2$ and the fast and diffuse J-shock traced by ionic lines are given in Tables \ref{tbl:shockmodels} and \ref{table:diagnostics}, respectively. The highest pressures are generally experienced in the molecular C-shocks in which the post-shock gas cools and condenses and can obtain a pressure higher than the thermal pressure of the remnant proportional to the gas compression factor \citep{moorhouse91}. It is well established theoretically that pressure enhancements of 10-100 times can occur when the SNR shell interacts with dense clumps \citep{chevalier99}. OH maser emission detected from Kes 69, G348.5-0.0, G346.6-0.2, and G349.7+0.2, which is evidence of the survival of higher density clumps. We caution that the pressures derived from shock models may have uncertainties up to an order of magnitude.  Given this large uncertainty it is still clear that H$_2$ emission originates from regions where the pressure is of order 10$^{-7}$--10$^{-6}$ dyn cm$^{-2}$. The pressures derived for each of the two C-shock components is consistent with an isobaric shock into a multi-phase cloud with a range of densities. 
In general the pressure derived from the three ionic lines are comparable to that derived for the C-shocks, though the densities are lower (100--1000 cm$^{-3}$) and shock velocities higher. Again this is consistent with the SNR shock interacting with a multi-phase MC with both an ambient molecular phase and a dense clump phase \citep{reach00}. 

There are a few exceptions to the general observations above. G349.7+0.2 has an exceptionally high ionic 
shock ram pressure, even exceeding that of the dense C-shocks. 
The bright ionic lines, inferred high shock velocity and a high electron density inferred from [Fe II] line
suggest that this SNR is younger than other SNRs in our sample, and may contain a contribution from SN ejecta. 
X-ray observations also show abundance enhancements indicative of a significant ejecta mass \citep{lazendic05}
and the esatimated age of the SNR is 3500 years.

Another exception is G346.6-0.2, where only very weak ionic lines and OPR of 3 are observed. 
The ionic pressure is estimated to be only $\sim$10$^{-8}$ dyn cm$^{-2}$ whereas the pressure from the C-shocks are both 4\x10$^{-6}$ dyn cm$^{-2}$. Given that the OPR is observed at equilibrium, this may suggest that the shock has had sufficient time to establish a steady-state C-shock, whereas those remnants with non-equilibrium OPR and observed ionic emission may have not yet obtained steady-state. The weak ionic lines appear to arise from a shock into a low density ambient medium of 100 cm$^{-3}$ and not the dense clump material from which H$_2$ emission originates.

Finally, 3C 396 shows a large range of pressures. The pressure of the hot thermal X-ray gas observed in the interior of the SNR is estimated to be of order 3\x10$^{-9}$ dyn cm$^{-2}$ \citep{harrus99}. This is a factor of a few smaller than the shock pressure derived from ionic line diagnostics 1.8\x10$^{-7}$ dyn cm$^{-2}$. In comparison to the observed H$_2$ emission from 3C396, the shock pressure is observed to be a factor of $\sim$10 higher. Given the spatial segregation of the ionic and H$_2$ lines, it is plausible that H$_2$ emission is arising 
from interaction with dense clumps while emission from the main ionic lines arises from the same shock propagating into a 
lower ambient density medium or contribution from ejecta.
The fact that Fe
line emitting gas is possibly hotter and denser than the other four SNRs (see Table \ref{table:diagnostics}) and the
two different shock velocities support that the infrared emission from 3C 396 is composed of both ejecta
fragments and shocked ISM. Full spectral mapping (such as of Cas A; Rho et al. 2008) would be required to separate distribution of 
ejetcta and  shocked ISM.

\section{Diffuse Warm H$_2$ \label{sec:diffuse}}

Warm diffuse H$_2$ gas is known to be abundant  throughout the galaxy \citep{gry02,falgarone05}.
Diffuse H$_2$ emission has also been noted at the SNR-cloud interaction site, though it's relation
to the shock interaction is unclear \citep{neufeld07}. Using careful local background subtraction we
are able to isolate the H$_2$ emission due only to the supernova remnant interaction. We find
enhanced S(0) and S(1) emission from the shock interaction, but often the diffuse emission is dominant. Here
we discuss the excitation of this warm diffuse molecular gas.

Diffuse S(0) and S(1) lines are detected for nearly every sight line, including our sky reference
positions. The diffuse component we observe could arise from conditions in the pre-shock clouds or
from low density diffuse gas present along the line of sight through the Galaxy. Several excitation
mechanisms for warm diffuse H$_2$ present in the Galactic plane have been proposed including the
Galactic radiation field, H$_2$ formation energy (Spitzer 1974), and MHD shocks or small-scale
turbulence (Falgarone et al. 2005). However, higher energy transitions such as S(2) through S(7) are
not detected as for the SNRs. Because we have only one ortho and one para line, the ortho-to-para
ratio and gas temperature are degenerate so we can only constrain the excitation of diffuse H$_2$.

\begin{figure} \centering
\includegraphics[width=3in,angle=0]{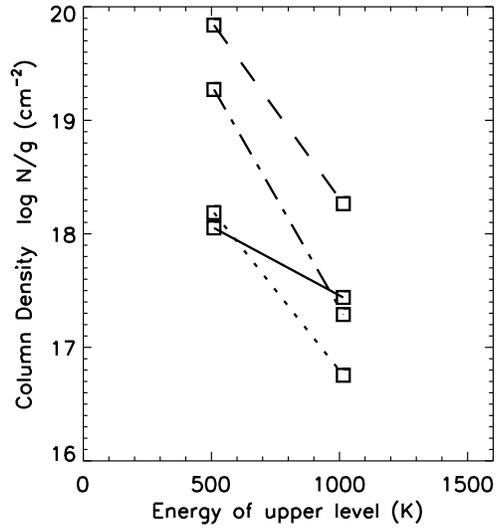}
\caption{Boltzmann diagram for the excitation of diffuse H$_2$ observed toward SNRs in our sample.
The SNRs are color-coded with blue for G346.6-0.2, red for 3C 396, purple for G349.7+0.2, and yellow
for G348.5-0.0.
\label{fig:h2compare}}
\end{figure}

Table \ref{tbl:diffusefits} gives the fitted parameters to the warm diffuse H$_2$ emission in the regions 
used as a local background in data reductions. 
We have not included Kes 69 or Kes 17 as the slit of one of the LL orders lies almost entirely inside 
the SNR. As the temperature is degenerate with the ortho-to-para ratio, we have assumed a range of different OPR. In the first block of physical parameters listed in Table 10 we assume an OPR of 3.0 to derive excitation temperatures that range from 100 to 160 K, with the notable exception of G346.6-0.2. 
Figure \ref{fig:h2compare} shows the Boltzmann diagram for diffuse emission from outside the SNRs. 
The slope of the line connecting the S(0) and S(1) transitions is markedly different for G346.6-0.2 
than for the other remnants. Diffuse emission towards G346.6-0.2 may be associated with  a foreground 
HII region which appears to dominate the MIPS 24\micron\ image of the SNR's vicinity. 

The assumption of an ortho-to-para ratio of three gives the minimum temperature of the gas (see Table 10, row 1). At such
low temperatures the equilibrium OPR is roughly two and may be a more appropriate value. However, if
the diffuse warm H$_2$ has only been recently heated from a colder, quiescent state it could have a
significantly lower OPR. For comparison with the shocked gas, we use the OPR of the warm shocked gas
listed in Table \ref{tbl:h2exfits}. These solutions for various ortho-to-para ratios are all given
in the second block of parameters listed in Table \ref{tbl:diffusefits}.

Even for the lowest OPR (the third block of parameters listed in Table 10), which is taken to be the same as that observed for the SNR shock, the
excitation temperature of the low-lying rotation transitions is comparable to the 180-390 K
excitation temperatures observed by Savage et al. (1977) in a study of UV absorption sight lines.
Falgarone et al. (2005) reports on the excitation of S(0) through S(3) transitions for a line of
sight devoid of massive star forming regions finding comparable surface brightness for the S(1) and
S(0) transitions. As we did not detect the S(2) through S(7) lines for the diffuse molecular gas we
were not able to probe these proposed signatures of dissipative turbulence.

Though the excitation conditions are somewhat uncertain, we are able to estimate the fraction
of warm molecular gas to the total amount of molecular gas. Using the CO survey of \citep{dame01} we
integrate the amount of CO along the line of sight towards the SNR. We estimate the total column of
molecular hydrogen using the canonical conversion N(H$_2$)/I$_{CO}$=2$\times$10$^{20}$ cm$^{-2}$
(K\kms )$^{-1}$. Though our apertures differ greatly, if the diffuse component is fairly uniformly
distributed we estimate a warm gas fraction of 0.005 and 0.15  towards SNRs G346.6-0.2 and
G348.5-0.0 respectively.

\section{Conclusions}
The nature of six SNRs identified as having molecular colors in the GLIMPSE survey have been
resolved spectrally by {\it Spitzer} IRS observations. Strong H$_2$ S(0) to S(7) lines show clear
evidence of shock interaction with dense gas. This doubles the number of known molecular remnants
for which H$_2$ emission has been characterized. The excitation of H $_2$ is well fit by warm
T$\sim$400 K and hot T$\sim$1500 K gas components. The OPR of the warm component deviates from
equilibrium consistent with the passage of a slow shock through dense gas.

Ionic fine structure lines are also present for all SNRs but have large variations in brightness
between the remnants. The presence of both strong molecular emission and ionic fine structure lines
with large ionization potentials requires multiple shocks present within the observed aperture. In
3C 396 we see that the H$_2$ and ionic lines are spatially separated, indicating changing physical
conditions across the shock front which agrees with a multi-phase pre-shock medium composed of
atomic, molecular and dense clump gas.

Comparison of shock models shows that two C-shocks are required to explain the observed H$_2$
emission. Fitted shock parameters show one shock is of low velocity ($\sim$10\kms ) into dense gas
($\sim$10$^6$ cm$^{-3}$) while another  shock of higher velocity ($\sim$50\kms ) is propagating into
lower density gas ($\sim$10$^4$ cm$^{-3}$). Ionic lines require a fast J-shock into diffuse gas
typically with a density $\sim$10$^3$ cm$^{-3}$. We find physical conditions for these fast ionic
shocks imply pressures similar to those found in the molecular C- shocks into dense clumps. For
G349.7+0.2 we find extremely high ionic line pressures which are several times higher than for the
H$ _2$ emitting gas, possibly indicative of powerful ejecta shocks.

Long slit spectroscopy of these SNRs isolates diffuse emission from that of the remnant. This is
particularly important for the H$_2$ S(0) and S(1) lines as well as diffuse interstellar cooling
lines such as [Si II] 34.8\micron . We find the diffuse H$_2$ gas is relatively cool with a
temperature of 100--300 K and a warm gas fraction of 0.5--15 \% .

\bigskip
\begin{acknowledgments}
We thank Achim Tappe for updating atomic data and fruiteful discussion on atomic Fe lines.  
This work is based on observations made with the {\it Spitzer Space Telescope}, which is 
operated by the Jet Propulsion Laboratory under NASA contract. J.W.H. is grateful for the 
support of  the IPAC staff and the Spitzer Visiting Graduate Student Program.
\end{acknowledgments}

\begin{deluxetable}{lllllllll}
\tablecaption{Properties of Supernova Remnants\label{tbl:properties}}
 \tabletypesize{\scriptsize}
 \tablewidth{0pt}
 \tablehead{& \colhead{Kes 69} & \colhead{3C 396} & \colhead{Kes 17} & \colhead{G346.6-0.2} & \colhead{G348.5-0.0} & \colhead{G349.7+0.2} }
 \startdata
%               			& Kes69    & 3C 396    & Kes17    & G346.6	   & G348.5   & G349.7   
RA(J2000)		&18:33:01.89 &19:03:56.21 &13:05:32.75 &17:13:43.82 &17:15:04.90 &17:21:24.90 &\\
DEC(J2000)		&-10:13:44.2 &+05:25:49.7 &-62:40:06.0 &-40:18:06.0 &-38:33:41.0 &-37:29:11.6 &\\
diameter ($'$)     		& 20       & 8\x6     & 8        & 8        & 10       & 2.5      &\\
distance (kpc)     		& 5.2      & $\ge$7  &$\ge$9.7  & 11       & 13.7     & 22       &\\
physical size (pc) 		& 30       & $\ge$16\x12 &$\ge$23 & 26       & 39       & 16       &\\
velocity (km s$^{-1}$)	& +80,+69  & ?        & ?        & --76     & --22     & +16      &\\
%age (yrs)           		& ?        & 7000     &  ?       & ?        & ?        & 2800     &\\
N$_H$ (10$^{22}$ cm$^{-2}$)& 2.4 & 4.7 & 3.6 & 2.68%$^{+2.32}_{-0.68}$ 
& 2.8 & 5.8 &\\
$\tau$(9.7)	& 0.95     & 1.8      & 1.4      & 1.05     & 1.1      & 2.3      &\\
A$_V$ (mag)				& 13       & 25       & 19.3     & 14.3     & 15       & 31       &\\
References$^{a}$ 		& 1,2      & 3,4      & 5,11    & 6,7,11   & 5,10,11 & 8,9      &
\enddata
\tablenotetext{a}{References (1)\citet{wilson72}, (2)\citet{fyz03mmsnrs}, (3)\citet{harrus99}, (4)\citet{olbert03}, (5) \citet{caswell75},(6)\citet{dubner93}, (7)\citet{koralesky98}, (8)\citet{slane02}, (9)\citet{lazendic05},  (10) \citet{reynoso00} (11) this work.}
\end{deluxetable}

\begin{deluxetable}{ll|cccccc}
\tabletypesize{\scriptsize}
\tablewidth{0pt}
\tablecaption{Observed Brightness of Emission Lines After Local Background Subtraction\label{tbl:observed}}
\tablehead{
& & \multicolumn{6}{c}{Intensity (ergs cm$^{-2}$ s$^{-1}$ sr$^{-1}$)}  \\
\colhead{Transition} & \colhead{$\lambda$($\mu$m)}& \colhead{Kes 69} & \colhead{3C 396} 
& \colhead{Kes 17} & \colhead{G346.6-0.2} & \colhead{G348.5-0.0} & \colhead{G349.7+0.2}
}
\startdata
%&&($\mu$m)         &&Kes 69        &3C 396          &Kes17          &G346.6-0.2		&G348.5-0.0		&G349.7+0.2	&\\ \hline
H$_2$ S(0) & 28.219 &4.29(0.28)\e-6 &6.85(2.90)\e-6 &4.54(0.28)\e-6 &1.34(0.25)\e-6 &3.26(0.27)\e-6 &1.54(0.15)\e-5\\
H$_2$ S(1) & 17.035 &4.14(0.03)\e-5 &3.15(0.41)\e-5 &1.00(0.01)\e-4 &4.26(0.05)\e-5 &1.27(0.04)\e-5 &8.18(0.10)\e-5\\
H$_2$ S(2) & 12.279 &4.51(0.02)\e-5 &4.02(0.41)\e-5 &1.58(0.30)\e-4 &4.60(0.07)\e-5 &2.69(0.07)\e-5 &1.40(0.02)\e-4\\
H$_2$ S(3) &  9.665 &8.44(0.24)\e-5 &2.56(0.19)\e-5 &2.20(0.24)\e-4 &1.32(0.02)\e-4 &2.89(0.11)\e-5 &6.89(0.17)\e-5\\
H$_2$ S(4) &  8.025 &1.24(0.16)\e-4 &1.14(0.12)\e-4 &4.56(0.55)\e-4 &1.20(0.05)\e-4 &8.76(0.32)\e-4 &2.32(0.05)\e-4\\
H$_2$ S(5) &  6.910 &2.03(0.10)\e-4 &3.29(0.21)\e-4 &1.28(0.09)\e-3 &3.64(0.19)\e-4 &2.99(0.04)\e-4 &4.61(0.07)\e-4\\
H$_2$ S(6) &  6.109 &1.06(0.05)\e-4 &6.15(0.78)\e-5 &3.80(0.32)\e-4 &6.45(0.56)\e-5 &8.87(0.41)\e-5 &---$^1$ \\
H$_2$ S(7) &  5.511 &3.15(0.09)\e-4 &1.60(0.14)\e-4 &7.56(0.07)\e-3 &2.22(0.05)\e-4 &2.47(0.05)\e-4 &3.20(0.10)\e-4\\
\hline
$[$Fe II]  &  5.35  &1.27(0.01)\e-4 &2.82(0.11)\e-4 &1.55(0.07)\e-4 &---            &8.55(0.07)\e-4 &1.49(0.02)\e-3 \\
$[$Ar II]  &  6.98  &---            &---            &---            &---            &2.05(0.05)\e-4 &1.58(0.01)\e-3 \\
$[$Ar III] &  9.00  &---            &---            &---            &---            &---            &2.35(0.41)\e-5 \\
$[$Ne II]  & 12.8   &1.50(0.01)\e-4 &1.10(0.08)\e-4 &1.18(0.03)\e-4 &3.20(0.08)\e-5 &3.50(0.01)\e-4 &2.92(0.02)\e-3 \\
$[$Ne III] & 15.5   &3.72(0.24)\e-5 &1.51(0.08)\e-4 &4.43(0.02)\e-5 &2.47(0.28)\e-6 &7.42(0.03)\e-5 &1.11(0.01)\e-3 \\
$[$Fe II]  & 17.9   &8.37(0.27)\e-6 &3.55(0.04)\e-5 &1.17(0.06)\e-5 &---            &5.88(0.05)\e-5 &3.77(0.02)\e-4 \\
$[$S III]  & 18.7   &4.19(0.26)\e-6 &3.65(0.03)\e-5 &8.83(0.34)\e-6 &1.62(0.26)\e-6 &1.59(0.03)\e-5 &2.43(0.03)\e-4 \\
$[$Fe II]  & 24.5   &3.44(0.23)\e-6 &1.31(0.38)\e-5 &2.76(0.13)\e-6 &---            &1.62(0.03)\e-5 &1.90(0.03)\e-4 \\
$[$S I]    & 25.2   &5.64(0.29)\e-6 &---            &7.94(0.25)\e-6 &1.95(0.30)\e-6 &1.77(0.02)\e-5 &1.17(0.02)\e-4 \\
$[$Fe II]$^2$& 25.9 &3.84(0.02)\e-5 &9.94(0.01)\e-5 &4.54(0.01)\e-5 &7.00(0.19)\e-6 &1.80(0.01)\e-4 &9.82(0.05)\e-4 \\
$[$S III]  & 33.5   &1.52(0.04)\e-5 &5.58(0.09)\e-5 &1.94(0.03)\e-5 &3.45(0.30)\e-6 &3.35(0.03)\e-5 &4.02(0.03)\e-4 \\
$[$Si II]  & 34.8   &2.53(0.01)\e-4 &2.12(0.05)\e-4 &2.30(0.01)\e-4 &6.93(0.05)\e-5 &7.14(0.02)\e-4 &2.12(0.02)\e-3 \\
\enddata
\\$^1$ Blended with strong PAH feature. $^2$ Blended with $[$O IV] 25.9\micron\ line.
\end{deluxetable}

\begin{deluxetable}{ll|cccccc}
\tablecaption{De-reddened brightness of emission lines assuming an extinction curve of R$\mathrm{_V}=3.1$. }
\tabletypesize{\scriptsize}
\tablewidth{0pt}

\tablehead{
& & \multicolumn{6}{c}{Intensity (ergs cm$^{-2}$ s$^{-1}$ sr$^{-1}$)}  \\
\colhead{Transition} & \colhead{$\lambda$($\mu$m)}& \colhead{Kes 69} & \colhead{3C 396} 
& \colhead{Kes 17} & \colhead{G346.6-0.2} & \colhead{G348.5-0.0} & \colhead{G349.7+0.2}
}
\startdata
%&&($\mu$m)         &&Kes 69        &3C 396          &Kes17          &G346.6-0.2		&G348.5-0.0		&G349.7+0.2	&\\ \hline
H$_2$ S(0) & 28.219 &4.58(1.51)\e-6 &9.69(4.10)\e-6 &1.33(0.22)\e-5 &1.49(0.30)\e-6 &4.01(0.27)\e-6 &2.11(0.19)\e-5\\
H$_2$ S(1) & 17.035 &5.50(0.06)\e-5 &5.70(0.75)\e-5 &1.98(0.13)\e-4 &6.01(0.16)\e-5 &1.81(0.44)\e-5 &1.88(0.39)\e-4\\
H$_2$ S(2) & 12.279 &5.75(0.07)\e-5 &8.03(0.81)\e-5 &2.68(0.17)\e-4 &6.83(0.11)\e-5 &4.07(0.11)\e-5 &3.92(0.66)\e-4\\
H$_2$ S(3) &  9.665 &1.45(0.08)\e-4 &1.47(0.11)\e-4 &9.09(0.50)\e-4 &3.44(0.38)\e-4 &8.18(0.30)\e-5 &5.40(0.17)\e-4\\
H$_2$ S(4) &  8.025 &1.56(0.14)\e-4 &2.15(0.23)\e-4 &7.59(0.55)\e-4 &1.75(0.07)\e-4 &1.28(0.04)\e-4 &3.53(0.68)\e-4\\
H$_2$ S(5) &  6.910 &3.49(0.28)\e-4 &4.61(0.30)\e-4 &1.71(0.09)\e-3 &3.58(0.86)\e-4 &3.66(0.05)\e-4 &5.19(1.06)\e-4\\
H$_2$ S(6) &  6.109 &1.13(0.12)\e-4 &9.36(1.19)\e-5 &5.10(0.32)\e-4 &8.36(0.66)\e-5 &1.14(0.05)\e-4 &---$^1$ \\      
H$_2$ S(7) &  5.511 &3.44(0.21)\e-4 &2.48(0.21)\e-4 &1.00(0.07)\e-3 &2.78(0.09)\e-4 &3.21(0.06)\e-4 &5.34(0.62)\e-4\\
\hline
$[$Fe II]  &  5.35  &1.61(0.10)\e-4 &4.47(0.07)\e-4 &2.22(0.10)\e-4 &---            &1.13(0.02)\e-3 &2.50(0.02)\e-3 \\
$[$Ar II]  &  6.98  &---            &---            &---            &---            &2.51(0.05)\e-4 &2.30(0.01)\e-3 \\
$[$Ar III] &  9.00  &---            &---            &---            &---            &---            &1.19(0.20)\e-4 \\
$[$Ne II]  & 12.8   &2.01(0.09)\e-4 &1.96(0.15)\e-4 &1.83(0.04)\e-4 &4.44(0.09)\e-5 &4.93(0.01)\e-4 &5.51(0.03)\e-3 \\
$[$Ne III] & 15.5   &4.74(0.04)\e-5 &2.43(0.13)\e-4 &6.37(0.03)\e-5 &3.24(0.36)\e-6 &9.84(0.04)\e-5 &1.88(0.01)\e-3 \\
$[$Fe II]  & 17.9   &1.16(0.04)\e-5 &6.71(0.06)\e-5 &1.90(0.10)\e-5 &---            &8.59(0.08)\e-5 &7.62(0.02)\e-4 \\
$[$S III]  & 18.7   &5.79(0.36)\e-6 &6.88(0.05)\e-5 &1.44(0.06)\e-5 &2.33(0.38)\e-6 &2.32(0.04)\e-5 &4.90(0.03)\e-4 \\
$[$Fe II]  & 24.5   &4.23(0.29)\e-6 &1.97(0.38)\e-5 &3.76(0.17)\e-6 &---            &2.06(0.03)\e-5 &2.97(0.03)\e-4 \\
$[$S I]    & 25.2   &6.88(0.36)\e-6 &---            &1.07(0.03)\e-5 &2.44(0.32)\e-6 &2.23(0.03)\e-5 &1.80(0.02)\e-4 \\
$[$Fe II]$^2$& 25.9 &4.66(0.03)\e-5 &1.45(0.02)\e-4 &6.06(0.02)\e-5 &8.68(0.21)\e-6 &2.25(0.02)\e-4 &1.49(0.01)\e-3 \\
$[$S III]  & 33.5   &1.76(0.05)\e-5 &7.42(0.19)\e-5 &2.41(0.03)\e-5 &4.06(0.30)\e-6 &3.97(0.04)\e-5 &5.50(0.04)\e-4 \\
$[$Si II]  & 34.8   &2.90(0.04)\e-4 &2.76(0.07)\e-4 &2.82(0.01)\e-4 &8.06(0.05)\e-5 &8.36(0.03)\e-4 &2.84(0.04)\e-3 \\
\enddata
\\$^1$ Blended with strong PAH feature. $^2$ Blended with $[$O IV] 25.9\micron\ line.
\end{deluxetable}

\begin{deluxetable}{l|rrrrrr}
\tabletypesize{\scriptsize}
\tablewidth{0pt}
\tablecaption{Comparison of IR Line Luminosities}
\tablehead{
\colhead{Luminosity ($\lsol$) \label{tbl:lum}}
& \colhead{Kes 69}
& \colhead{3C 396}
& \colhead{Kes 17}
& \colhead{G346.6-0.2} 
& \colhead{G348.5-0.0} 
& \colhead{G349.7+0.2}}
\startdata
% & Kes 69 &3C 396   &Kes17    &G346.6-0.2&G348.5-0.0&G349.7+0.2&\\
H$_2$ S(0)-S(7) 
   & 1.4 & 2.5 & 21 & 7.2 & 6.8 & 68 \\ 
H$_2$ S(0)-S(2) 
   & 0.5 & 0.9 & 6.3 & 2.2 & 1.5 & 37 \\
H$_2$ S(5)-S(7) 
   & 0.6 & 0.9 & 8.2 & 2.3 & 3.9 & 13 \\
Ar & -- & -- & -- & -- & 1.4 & 36 \\
Fe & 0.7 & 3.8 & 3.2 & 0.4 & 26 & 448 \\
Ne & 0.6 & 2.6 & 2.3 & 0.4 & 9.5 & 358 \\
S  & 0.3 & 2.4 & 1.5 & 0.3 & 4.9 & 201 \\
Si & 4.4 & 6.5 & 15 & 5.4 & 85 & 764 \\
\enddata 
\end{deluxetable}

\begin{deluxetable}{l|rrrrrrr}
\tabletypesize{\scriptsize}
\tablewidth{0pt}
\tablecaption{Fitted Excitation Parameters to Observed Molecular Hydrogen Lines}
\tablehead{
& \colhead{Kes 69}
& \colhead{3C 396}
& \colhead{Kes 17}
& \colhead{G346.6-0.2} 
& \colhead{G348.5-0.0} 
& \colhead{G349.7+0.2}
}
\startdata
N(H$_2$)$_{warm}$ &1.5\e20 &3.0\e20 & 3.8\e20 & 4.7\e19 & 1.2\e20 & 2.8\e20 \\
T$_{warm}$ (K)    &303     &265     & 322     & 528     & 261     & 467 \\
OPR$_{warm}$      &1.3     &0.65    & 1.4     & 3.0     & 0.4     & 0.99 \\
\hline
N(H$_2$)$_{hot}$  &5.0\e18 &6.9\e18 & 2.6\e19 & 8.3\e18 & 2.4\e18 & 5.2\e18 \\
T$_{hot}$ (K)     &1465    &1270    & 1152    & 1241    & 1773    & 1647 \\
OPR$_{hot}$       &3.0     &3.0     & 2.5     & 3.0     & 3.0     & 3.0 \\
\enddata \label{tbl:h2exfits}\end{deluxetable}

\begin{deluxetable}{r|llllll} \tabletypesize{\scriptsize}
\setlength{\tabcolsep}{0.03in}
\tablewidth{0pt} \tablecaption{Combination of Two C-Shock Model Fits \label{tbl:shockmodels}}
\tablehead{& \colhead{Kes 69} & \colhead{3C 396} & \colhead{Kes 17} & \colhead{G346.6--0.2} & \colhead{G348.5--0.0} & \colhead{G349.7+0.2}}
\startdata
\underline{C-shock 1} \\
% Kes 69      &3C 396       &Kes17        &G346.6       &G348.5       &G349.7
n$_0$ (cm$^{-3}$)
 & 10$^5$      & 10$^6$      &10$^6$       & 10$^6$      & 10$^5$      &10$^6$ \\
v$_s$ (km s$^{-1}$) 
 & 10          & 10          & 10          & 15          & 10          &10 \\
$\Phi_s$     
 &0.043(4)     &0.089(2)     & 0.28(3)     &0.055(2)     &0.095(9)     &0.61(3) \\
OPR           
 & 1           & 2           & 1           & 3           & 1           &1 \\
P$_s$ (dyn cm$^{-2}$) 
 &2\x10$^{-7}$ &2\x10$^{-6}$ &2\x10$^{-6}$ &4\x10$^{-6}$ &2\x10$^{-7}$ &2\x10$^{-6}$ \\
N(H$_2$) (cm$^{-2}$)
 & 4.3\e20      & 5.4\e19    &2.0\e20 & 2.1\e19 & 1.0\e20 & 4.1\e20\\
% Kes 69      &3C 396       &Kes17        &G346.6       &G348.5       &G349.7
\hline
\underline{C-Shock 2} \\
n$_0$ (cm$^{-3}$)
 & 10$^4$      & 10$^5$      &10$^4$       & 10$^5$      & 10$^4$      &10$^5$ \\
v$_s$ (km s$^{-1}$) 
 & 40          & 30          & 40          & 50          & 40          &40 \\
$\Phi_s$     
 &0.029(1)     &0.0091(1)    & 0.094(1)     &0.004(1)     &0.024(1)     &0.0071(7) \\
P$_s$ (dyn cm$^{-2}$)
 &3\x10$^{-7}$ &1.5\x10$^{-6}$ &3\x10$^{-7}$ &4\x10$^{-6}$ &3\x10$^{-7}$ &3\x10$^{-6}$ \\
 %Kes 69       &3C 396       &Kes17        &G346.6       &G348.5       &G349.7
N(H$_2$) (cm$^{-2}$)
 & 1.5\e20 & 2.6\e18 & 5.0\e20 & 3.5\e24 & 1.3\e20 & 7.1\e19 \\
%\hline
%$\chi^2_r$ & 12    & 8         & 4.7         & 7         & 2.5    & 0.4 \\
\enddata 
\tablecomments{Errors to fitted shock models are found to be approximately
0.5 for gas density log$_{10} n$,  5\kms\ for shock velocity v$_s$, and less than 10\% for $\Phi_s$.}
\end{deluxetable}

\begin{deluxetable}{r|llllll} \tabletypesize{\scriptsize}
\setlength{\tabcolsep}{0.03in}
\tablewidth{0pt} \tablecaption{Combination of C- and J-Shock Model Fits \label{tbl:cjshockmodels}}
\tablehead{& \colhead{Kes 69} & \colhead{3C 396} & \colhead{Kes 17} & \colhead{G346.6--0.2} & \colhead{G348.5--0.0} & \colhead{G349.7+0.2}}
\startdata
%&            &&Kes 69 &3C 396   &Kes17    &G346.6 &G348.5 &G349.7
\underline{slow C-shock} \\
n$_0$ (cm$^{-3}$) 
              & 10$^6$ & 10$^6$ & 10$^6$& 10$^6$ & 10$^6$ & 10$^6$  \\
v$_s$ (\kms)  & 10     & 15     & 10    & 15     & 15     & 10 \\
$\Phi_s$      &0.132   &0.046   &0.13   &0.053   &0.064   &0.54 \\
OPR           & 2      & 2      & 2     & 3      & 1      & 1  \\ 
P$_s$ (dyn cm$^{-2}$)
 &2\x10$^{-6}$ &4\x10$^{-6}$ &2\x10$^{-6}$ &4\x10$^{-6}$ &4\x10$^{-6}$ &2\x10$^{-6}$ \\
%nv$_s^2$      & 1.0e9 & 1.0e8 & 1.0e8 & 1.0e9 & 1.0e8 & 1.0e9\\
%N(H$_2$) (cm$^{-2}$)&    &    &   & 2\e19   &    & \\
\hline
\underline{fast J-shock} \\
n$_0$ (cm$^{-3}$) & 10$^3$ & 10$^4$ & 10$^3$& 10$^4$       & 10$^3$ & 10$^4$ \\
v$_s$ (\kms)      & 90     & 130    & 50    & 150     & 60     & 110 \\
$\Phi_s$          &0.104&0.005&0.19&0.0053 &0.057 &0.10\\
P$_s$ (dyn cm$^{-2}$)
 &1.4\x10$^{-7}$ &3\x10$^{-6}$ &4\x10$^{-8}$ &3\x10$^{-6}$ &6\x10$^{-8}$ &2\x10$^{-7}$ \\
%\hline
%$\chi^2_r$ & 20 & 70 & 14.5 & 8.4         & 24   & 2.9 \\
\enddata 
\end{deluxetable}

\begin{deluxetable}{l|rrrrr}
\tablecaption{Comparison of model fitting for single and multiple shocks
\label{table:extra}}
 \tabletypesize{\scriptsize}
 \tablewidth{0pt}
 \tablehead{
Model & One C-shock  &  Two C-shocks & Two n$_0$=10$^4$ C-shocks  & C
\& J-shock  \\
SNR     & (n$_0$, v$_s$) [$\chi^2_r$] & (n$_0$, v$_s$),(n$_0$, v$_s$)
[ $\chi^2_r$ ] & (n$_0$, v$_s$),(n$_0$, v$_s$) [ $\chi^2_r$ ] &
(n$_0$, v$_s$),(n$_0$, v$_s$) [ $\chi^2_r$ ]}
\startdata
Kes 69      & (10$^5$,20)[60]& {\bf (10$^5$,10),(10$^4$,40)[12]} &
(10$^4$,20),(10$^4$,50)[24] & (10$^6$,10),(10$^3$,90)[20]  \\
3C 396      & (10$^6$,15)[56]& {\bf (10$^6$,10),(10$^5$,30)[8.0]}&
(10$^4$,20),(10$^4$,40)[23] & (10$^6$,15),(10$^4$,130)[65] \\ %
Kes 17      & (10$^5$,30)[23]& {\bf (10$^6$,10),(10$^4$,40)[4.1]}&
(10$^4$,20),(10$^4$,40)[5.4]& (10$^6$,10),(10$^3$,50)[48] \\ %
G346.6-0.2  & (10$^6$,15)[14]&      (10$^6$,15),(10$^5$,50)[7.0]&
(10$^4$,30),(10$^4$,50)[4.9] &  {\bf (10$^6$,15),(10$^4$,150)[0.9]} \\
G348.5-0.0  & (10$^6$,20)[15]& {\bf (10$^5$,10),(10$^4$,40)[2.4]}&
(10$^4$,10),(10$^4$,40)[4.4]& (10$^6$,10),(10$^3$,60)[12.4] \\ %
G349.7+0.2  & (10$^6$,15)[16]& {\bf (10$^6$,10),(10$^4$,50)[0.6]}&
(10$^4$,20),(10$^4$,50)[6.3]& (10$^6$,10),(10$^4$,110)[2.9]
\enddata
\\ {$^a$ The values in the brackets are reduced $\chi^2_r$. The best solution is in bold}
\end{deluxetable}

\begin{deluxetable}{l|rr|rr|r|r|r|}
\tablecaption{Physical properties derived from ionic line diagnostics \label{table:diagnostics}}
 \tabletypesize{\scriptsize}
 \tablewidth{0pt}
 \tablehead{
(source)$^a$  & \multicolumn{2}{c}{[Fe II] ratios}  & \multicolumn{2}{c}{[Si II], [Fe II], [Ne II] lines} & [Ne III]/[Ne II] & V$_s$ &Pressure$^b$\\
  SNR name & n$_e$ (cm$^{-3}$)  & T (K) &n$_o$ (cm$^{-3}$) & v$_s$ (km s$^{-1}$) & v$_s$ (km s$^{-1}$) & & (dyne cm$^{-2}$)}
 \startdata
Kes 69 &  \multicolumn{2}{c}{multi-T or n$_e$$^c$}  & 500-1000 & 150 & 150 & 150 &2.8$\times$10$^{-7}$\\ 
3C 396 &270 & 23000 &  500-1000     & 100-150 & 350 & 120,350 & 1.8$\times$10$^{-7}$,  1.5$\times$10$^{-6}$\\ 
Kes 17 & 100 & 7000   & 500-1000 &  150 & 150 or 200 & 150 &2.8$\times$10$^{-7}$\\
G346.6-0.2 & -- & -- & 100-200 & 70-100 & 80 & 80  &1.1$\times$10$^{-8}$ \\
G348.5-0.0 & \multicolumn{2}{c}{multi-T or n$_e$$^c$} & 500-1000 & 160 & 160 & 160 & 3.2$\times$10$^{-7}$ \\  
G349.7+0.2 &  700     &7000 & 10$^4$ & 250  & 80 or 250 &250 &1.0$\times$10$^{-5}$ \\ 
\enddata
\\ {$^a$ Denotes which ionic lines were used to derive the density, temperature or shock velocity.
$^b$ The pressure is estimated based on n$_o$ and V$_s$ derived from Si, Fe and Ne lines
and from [Ne III]/[Ne II] ratio. $^c$possible multi-tempeature and multi-density solution 
(see Fig. 8); the ratio of [Fe II] 17.9/5 $\mu$m yields a low density ($<$ 
400 cm$^{-3}$) and low ($<$2000 K)  temperature.}
\end{deluxetable}

\begin{deluxetable}{l|rrrrr} \tabletypesize{\scriptsize}
\tablewidth{0pt} \tablecaption{Fitted Physical Conditions of Diffuse H$_2$}
\tablehead{ & \colhead{3C 396} & \colhead{G346.6-0.2} & \colhead{G348.5-0.0} & \colhead{G349.7+0.2}}\startdata
N$_{H_2}$(cm$^{-2}$) 
         & 2.1e20 & 5.3e19 & 4.3e21 & 6.4e21 \\
T (K)    & 155     & 361     & 118     & 154 \\
OPR      & $\equiv$ 3.0     & $\equiv$ 3.0     & $\equiv$ 3.0     & $\equiv$ 3.0 \\
\hline
N$_{H_2}$(cm$^{-2}$) 
         & 1.2e20 & 3.6e19 & 2.40e21 & 3.6e21 \\
T (K)    & 177     & 509     & 130     & 176 \\
OPR      & $\equiv$ 2.0     & $\equiv$ 2.0     & $\equiv$ 2.0     & $\equiv$  2.0 \\
\hline
N$_{H_2}$(cm$^{-2}$) 
         & 3.3e19 &        & 3.7e21 & 1.5e21 \\
T (K)    & 292    &        & 222      & 233 \\
OPR       & 0.65   &        & 0.4     & 0.99 \\
\enddata \label{tbl:diffusefits} \end{deluxetable}

\vfill \newpage

\vfill \newpage

\vfill \newpage

\clearpage

\clearpage

\vfill \newpage

\end{document}